\documentclass[english]{article}
\usepackage[utf8]{inputenc}
\usepackage[T1]{fontenc}
\usepackage{amsmath}
\usepackage{graphicx}
\usepackage{fancyhdr}
\usepackage{float}
\usepackage{subcaption}
\usepackage{adjustbox}
\usepackage{multirow}
\usepackage{booktabs}
\usepackage{textgreek}
\usepackage{multirow}
\usepackage[dvipsnames]{xcolor}
\usepackage[hyphens]{url}
\usepackage{hyperref}
\usepackage{multirow}
\usepackage{tabularx}
\newcommand{\scidatalogo}{\includegraphics[height=36pt]{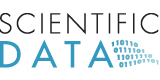}}
\newcommand{\overleaflogo}{\includegraphics[height=36pt]{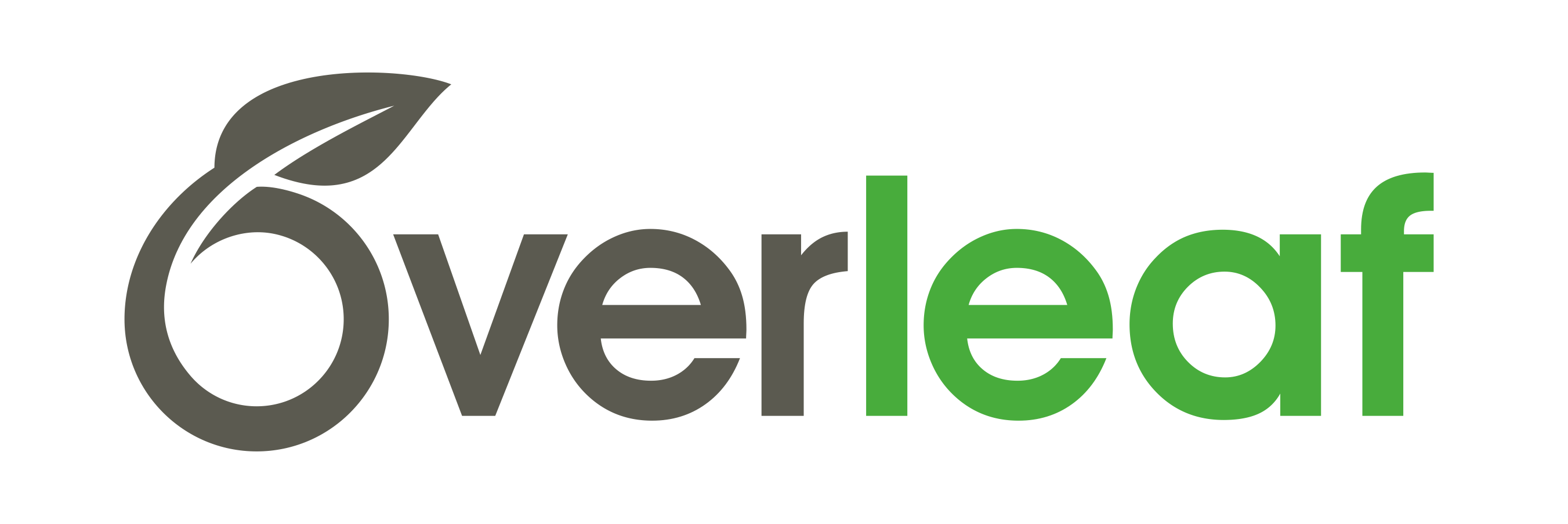}}
\begin{document}

\title{Citizen science dataset on residents' urban heat perception in outdoor public spaces of climate-vulnerable neighborhoods}

\author{Ferran Larroya\textsuperscript{1,2}, Isabelle Bonhoure\textsuperscript{1,2}, Femke Min\textsuperscript{1,2}, Josep Perell\'o\textsuperscript{1,2,{*}}}

\maketitle
\thispagestyle{fancy}

1. OpenSystems, Departament de F\'isica de la Mat\`eria Condensada, Universitat de Barcelona, Martí i Franquès, 1, 08028 Barcelona, Catalonia, Spain 

2. Universitat de Barcelona Institute of Complex Systems, Universitat de Barcelona, Barcelona, Catalonia, Spain

{*}corresponding author(s):
Josep Perell\'o (josep.perello@ub.edu)

\begin{abstract}
We present a dataset generated to investigate urban heat and thermal perception across five neighborhoods in the Barcelona metropolitan area. In collaboration with 14 non-academic partner organizations, we conducted a series of citizen science campaigns involving 439 residents as co-researchers engaged throughout all stages of the research process. Participants—residents of areas classified as highly or very highly climate-vulnerable—identified 210 public outdoor sites relevant to their daily lives. These locations were subsequently characterized using a range of spatial and environmental indicators pertinent to urban heat island effects, urban health, and climate resilience. Over the course of 48 thermal walks, participants carried portable, low-cost sensors that continuously recorded air temperature, relative humidity, and geolocation, resulting in 296,286 processed microclimatic data points. At pre-defined sites, individuals completed standardized surveys to report their Thermal Sensation Votes and Thermal Comfort Votes, yielding 5,169 self-reported entries. Sociodemographic data were also collected to further contextualize participants’ responses. The resulting dataset integrates objective environmental measurements with subjective perceptions of heat, enabling point-by-point analysis of thermal experience within the urban fabric. It offers a novel, multi-dimensional resource to support research on heat, thermal inequality, and the experiential dimensions of climate vulnerability, and is intended to inform evidence-based decision-making in urban planning, public health, and climate adaptation.
\end{abstract}

\section*{Background \& Summary}

Rising global temperatures and more frequent heat waves are among the most significant consequences of climate change \cite{IPCC_2021,Domeisen_2023}. Urban environments are particularly affected by high temperatures due to the urban heat island effect \cite{Ward_2016,Zhao_2018}. The combination of dense built environments, limited vegetation, and high energy consumption elevates ambient temperatures in urban areas \cite{Jayasooriya_2024}. Public outdoor spaces play a critical role in shaping both thermal conditions and residents’ capacity to cope with heat: they can either magnify heat stress through lack of shade and extensive hard surfaces or help mitigate it by offering cooling features and refuge \cite{Foshag_2020,Ahmed_2024,Yang_2025}. 

Extreme urban heat conditions exacerbate health risks \cite{Yang_2024}, particularly among vulnerable populations such as older adults, children, and individuals with pre-existing medical conditions \cite{Chen_2023, Monteiro_2024}. These risks compound existing inequities in the spatial distribution of heat exposure and its consequences \cite{Hsu_2021, Mashhoodi_2021, Guardaro_2022}. Marginalized populations—characterized by significant racial and income disparities \cite{Hsu_2021}, as well as groups such as the elderly and women \cite{Mashhoodi_2021} —experience disproportionate exposure to elevated urban heat intensity and face reduced adaptive capacity. This intersection of socioeconomic, gender, and age-related factors illustrates the multifaceted nature of heat inequities \cite{Guardaro_2022}. Comprehensive datasets that capture socio-demographic disparities in heat exposure—together with detailed information on the urban fabric—are essential for supporting equitable public health policies and data-driven climate adaptation strategies in urban environments \cite{Lobo_2023,Jancewicz_2024,Lu_2024}.

Recent research highlights the value of citizen science approaches for collecting environmental data in situ, at scales and resolutions difficult to achieve with conventional monitoring networks \cite{Fraisl_2022,Perello_2021,Nachtlichter_2025,DeRamos_2025}. Recent systematic literature reviews critically examine the evolving role of citizen science in advancing urban sustainability, planning, and transformative action \cite{Bonhoure_2025,Beck_2025}. Studies in this context highlight how citizen-generated data and participatory research approaches can inform policy, foster inclusive decision-making, and support the co-creation of more resilient and equitable urban environments \cite{VanBrussel_2019,Mahajan_2022,Pitidis_2024}.

Citizen science is taking a growing role in understanding and addressing urban heat phenomena \cite{Bonhoure_2025,Beck_2025}. Citizen-generated data have proven valuable for detecting and mapping urban heat (and cool) islands and heat exposure with high spatial and temporal resolution \cite{Rajagopalan_2020,Venter_2021,Rosso_2022,Kisters_2022,Wang_2023,Calhoun_2024}. Such approaches complement satellite and model-based measurements, offering hyperlocal insights that are often inaccessible through conventional monitoring systems \cite{Venter_2021,Chapman_2016,Lam_2021,Leichtle_2023}. Research points to the value of citizen science in revealing ecological responses to urban heat and in characterizing both hot and cool urban zones across cities \cite{Alonso_2019,Lehnert_2021,Dzyuban_2022,Best_2023,Lehnert_2023}. In addition to data generation, urban heat citizen science initiatives enhance public engagement and awareness of climate challenges, fostering co-learning and action-oriented strategies for heat adaptation \cite{Rajagopalan_2020,Kim_2024}. 

Originating in the field of urban studies, {\it thermal walks} investigate pedestrian thermal perception by integrating analyses of urban microclimate and spatial morphology \cite{Vasilikou_2020}. We argue that this methodology holds significant promise within the framework of citizen science, particularly in its capacity to engage individuals directly in the co-production of climate knowledge. Walking is conceptualized as a dynamic and situated practice that foregrounds the experiential diversity of urban environments \cite{Vasilikou_2018}. The methodology combines environmental monitoring with real-time subjective surveys, enabling a dual exploration of objective climatic conditions and lived thermal experiences \cite{Vasilikou_2020}. 

Thermal walks involve point-to-point assessments of thermo-spatial variation, capturing how individuals perceive and respond to heat while moving through the urban fabric \cite{Peng_2022}. By aligning microclimatic and spatial data with individuals’ perceptions, thermal walks facilitate a fine-grained understanding of the interplay between built form, environmental conditions, and thermal comfort \cite{Vasilikou_2020}. In recent years, this methodology has been further developed and adapted across diverse urban settings and socio-demographic groups, serving as a valuable tool to investigate heat exposure, thermal inequality, and the experiential dimensions of climate vulnerability \cite{Dzyuban_2022,Xie_2022,Smail_2024,Kruger_2025}.

The current Data Descriptor paper therefore presents a dataset generated through the Heat Chronicles citizen science campaigns conducted in the Barcelona metropolitan area. This region, situated in Catalonia, Spain, is classified as having a Mediterranean climate (Csa) according to the Köppen climate classification system \cite{Peel_2007}. However, the accelerating impacts of climate change have raised questions about whether this classification remains adequate for capturing emerging climatic realities in the region \cite{Beck_2018}. 

The Heat Chronicles implemented a series of thermal walks aiming to incorporate public participation at all stages of the research process. These campaigns can be situated within the framework of citizen social science, which highlights the active role of co-researchers—residents who possess local, experiential knowledge of their neighborhoods and who share a collective concern about urban heat—in shaping both the research questions and outcomes \cite{Albert_2021,Bonhoure_2023}. 
Custom-designed materials were developed to support the data collection protocol, to enhance participant engagement and transform the process into an enjoyable experience. These materials also served to increase the campaign’s visibility within the neighborhood context.

The campaigns were designed to investigate urban heat and perception across several neighborhoods identified as having high or very high climate vulnerability \cite{Garcia_2022}. Residents were recruited from diverse communities within these areas, in collaboration with 14 partner organizations. The individuals belonged to social groups identified in the literature as particularly vulnerable to extreme urban heat \cite{Yang_2024,Chen_2023,Hsu_2021,Mashhoodi_2021}. For each thermal walk, groups were composed to be socially homogeneous (e.g., similar age range, similar socio-demographic profile), enabling targeted comparative analyses. Prior to data collection, participants attended structured preparatory workshops to collaboratively identify outdoor public sites of daily relevance for inclusion in the thermal walk protocol. Selected sites were categorized by public use, emphasizing their role in shaping everyday thermal experiences. Following data collection, participants discovered data from the the selected sites and the involved routes to explore and interpret the results collectively, promoting both awareness and action. Finally, key findings were shared with neighborhood residents, local organizations, and policy makers through diverse formats and channels, reinforcing the community-driven dimension of the research. 

Thermal walks followed pre-defined routes during which individuals carried portable MeteoTracker sensors that continuously recorded air temperature, relative humidity, and GPS coordinates—whose accuracy has been validated \cite{Barbano_2024,Loglisci_2024}. At designated sites along each route, individuals completed structured surveys designed to capture their perceptions of heat, thermal comfort, and contextual factors. These surveys were adapted from standardized thermal perception instruments. Specifically, we collected Thermal Sensation Votes (TSVs) and Thermal Comfort Votes (TCVs), which reflect subjective assessments of temperature and comfort, respectively. 

Thermal sensation is a well-established parameter in thermal comfort research \cite{Velt_2017}, formally defined in ISO 7730 \cite{ISO_7730} and ASHRAE Standard 55 \cite{ASHRAE_55}. In temperate climates, it is recommended to use a $7$-point scale ranging from "cold" to "hot" \cite{Schweiker_2016}, while 9-point scales with added "very cold" and "very hot" categories are recommended for extreme climate conditions \cite{Dzyuban_2022}. To better capture extreme heat events—frequently encountered during the Heat Chronicles campaigns—we adapted the original 9-point thermal sensation scale by removing the “cold” and “very cold” categories from the response ballots. 

In contrast, thermal comfort is defined by ASHRAE Standard 55 \cite{ASHRAE_55} as “that condition of mind which expresses satisfaction with the thermal environment.” It reflects a broader psychological construct that integrates, but is not restricted to, thermal sensation. TCVs were recorded using a 7-point scale ranging from “very uncomfortable” to “very comfortable,” following established methodologies in thermal comfort research \cite{Dzyuban_2022,Velt_2017}. We extended standard thermal comfort instruments by adding an item that asked respondents to rate their thermal comfort while walking from the previous site to the current one. Similar walking-focused field protocols have demonstrated sensitivity to thermo-spatial variation and microclimatic transitions in urban routes \cite{Li_2023}. To ensure data quality and support, each walk was accompanied by a trained member of the research team, who provided instruction on sensor use and facilitated the administration of surveys.

The dataset compiled through the Heat Chronicles citizen science campaigns provides a unique open-access resource on human heat perception. The dataset is highly diverse, both in terms of the neighborhoods included—all classified as highly or very highly climate-vulnerable—and in the sociodemographic profiles of participants, which range from children aged 9–11 to older adults. It includes contributions from 439 participants across 48 distinct walking trajectories, covering over 300 kilometers within the Barcelona metropolitan area. In total, the campaigns generated 448\,654 geolocated microclimatic records using portable sensors, resulting in 296\,286 processed records. These are paired with 1\,867 Thermal Sensation Votes (TSVs), 1\,867 Thermal Comfort Votes (TCVs), and 1\,435 Thermal Comfort Votes while walking (wTCVs), totaling 5\,169 perceptual votes. The dataset also includes 210 geolocated points of interest and the specific transects that connect them. Additionally, participants provided sociodemographic information, including age, gender, typical time spent in public space, and neighborhood familiarity.

The 210 geolocated outdoor public sites included in the dataset were each categorized according to their public function within two broad domains: {\it Basic Needs} and {\it Well-being and Free Time}. These sites span diverse public typologies—such as streets, squares, parks, playgrounds, and green areas—alongside areas surrounding key public infrastructure, including transport hubs, schools, primary care centers, libraries, and cultural venues. For each site, additional spatial and environmental descriptors were gathered, including surface type, Normalized Difference Vegetation Index (NDVI), and Sky View Factor (SVF), all of which are established determinants of urban heat island intensity and thermal perception \cite{Gascon_2016,Miao_2020}. Furthermore, climate adaptation and mitigation indicators—such as distance to the nearest public fountain, green space, or designated climate shelter—were incorporated to augment the dataset’s relevance for urban health and climate resilience research. In addition, the dataset also includes for each site the Index of Vulnerability to Climate Change (IVAC, from the Catalan acronym, a composite index developed for the Barcelona metropolitan region assessing environmental and social vulnerability to climate change at neighborhood scale) \cite{Garcia_2022}. 

The resulting dataset offers a multi-layered perspective by integrating high-resolution environmental sensor data, geospatial context, subjective assessments of heat perception, and vulnerability indicators, thereby enabling nuanced analyses of thermal comfort and equity in urban environments. By combining low-cost, easy-to-use sensors with structured surveys and community engagement, this work contributes to ongoing discussions on how heat vulnerability is defined and measured \cite{Yang_2024,Chen_2023,Monteiro_2024,Hsu_2021,Mashhoodi_2021}, particularly with respect to public space \cite{Jayasooriya_2024,Foshag_2020,Ahmed_2024}. It also highlights the role of citizen social science \cite{Albert_2021,Bonhoure_2023} in advancing more inclusive approaches to climate adaptation and environmental justice \cite{Guardaro_2022}. The dataset enables analysis of spatial variability in urban microclimates and their relationship to perceived thermal discomfort \cite{Vasilikou_2020}. Furthermore, the publicly available data support investigations into how heat vulnerability varies across socio-demographic groups and public space typologies, providing an evidence base for equitable public health policies and climate adaptation strategies in urban settings. By sharing data and protocols, this paper also encourages the development of open participatory urban planning practices that foreground residents’ lived experiences in identifying priority areas for cooling interventions. Therefore, the dataset shared offers a timely and valuable resource for the growing body of research on urban outdoor thermal comfort, providing actionable insights to support equitable climate adaptation in vulnerable urban contexts \cite{Lam_2024,Kvetonova_2024,Li_2024,Mahia_2025,Subedi_2025,Zhou_2025,Li_2025}.
 
\section*{Methods}

The Universitat de Barcelona Ethics Committee (IRB00003099) approved the Heat Chronicles citizen science campaigns. For the participation of primary and secondary schools, additional approval was obtained from the Catalonia Education Department. All participants aged 18 years or older provided written informed consent prior to data collection. For participants under 18 years of age, consent was obtained from a parent or legal guardian following established procedures.

No personal data were collected. Each participant was assigned a random identification number. The sociodemographic information collected (e.g., gender, age range) does not permit individual identification. No privacy concerns were identified that would preclude the public release of the processed data. GPS trajectories of the thermal walks do not contain information about participants’ home locations, as all routes originated from designated public spaces, typically squares or parks.

\subsection*{Locations, community partners and participants} 

The participation of local communities was a central component of the research design for the Heat Chronicles campaigns. In the Barcelona metropolitan area, neighborhoods particularly vulnerable to extreme heat were identified using publicly available data sources, including the Index of Vulnerability to Climate Change (IVAC) \cite{Garcia_2022}. 

Based on these and other indicators of heat vulnerability, several potential partner institutions were contacted in spring 2023 to establish collaborations. This initial outreach enabled a pilot study conducted in summer 2023, implemented in four locations in partnership with a public library, a primary school, a secondary school, and a civil society organization (CSO). The pilot phase served to refine the scientific methodology, adjust experimental protocols, and evaluate the overall feasibility of the project. Building on the initial experience from summer 2023, the network of partners was expanded in 2024 to include five locations (cities or neighborhoods within the Barcelona metropolitan area; see Table \ref{tab:city_neighborhood}). 

For consistency, each location is referred to by the abbreviations provided in Table \ref{tab:city_neighborhood} throughout the text, tables, and figures of this manuscript. The neighborhoods or small cities of SP, CI, CT, and MR all have an IVAC greater than 61, placing them among the most climate-vulnerable locations in the Barcelona metropolitan area \cite{Garcia_2022}. Despite this shared vulnerability, these locations differ significantly in population size and density. Population figures currently ranged from around 14\,000 residents in CI (a neighborhood within Barcelona) to over 30\,000 in the small city of MR. According to 2021 data, CT and MR were the most densely populated areas, with 497.7 and 411.4 inhabitants per hectare, respectively, while SV was the most spatially dispersed, with only 31.2 inhabitants per hectare. 

Although SV scores an IVAC index slightly lower, it would also rank among the most vulnerable areas if only socio-economic factors were considered \cite{Garcia_2022}. Average income per person ranges from 12\,304 EUR in CT to 15,870 EUR in CI—both below Barcelona’s citywide average of 17\,350 EUR. Access to public green space is highly uneven, with SV offering the most (20.28 m$^2$ per inhabitant) and CT the least (2.18 m$^2$). Educational attainment further underscores structural inequality. In MR, 72.4\% of the population has completed only primary or lower secondary education, compared to just 26.9\% in SP—a gentrified neighborhood experiencing significant tourism pressure. Taken together, these indicators reveal sharp spatial, socioeconomic, and educational contrasts. These contextual data are critical for designing equitable, data-driven urban climate adaptation strategies at the local level.

A total of 14 institutions, representing a range of organizational types and spanning the five municipalities, actively participated in the campaigns. These institutions are reported in Table~\ref{tab:partners}. Diverse participant groups, varying in age and sociodemographic characteristics, engaged in the campaigns. Particular attention was given to including groups in more vulnerable situations, such as children, adolescents, and older adults, as well as individuals who regularly spend extended periods in public spaces (e.g., toddlers' families, summer camp participants). Participant characteristics are summarized in Table~\ref{tab:city_participant}.

Table~\ref{tab:sociodem_all} presents the sociodemographic profile of the 439 unique participants who took part in the thermal walks. The profile includes gender identity, age range, time spent in outdoor public spaces, and self-reported knowledge of their neighborhood. These participants contributed a total of 1,867 TSVs, 1\,867 TCVs, and 1\,435 wTCVs—resulting in 5\,169 self-reported entries collected across 210 public outdoor sites. Tables~\ref{tab:sociodem_SP}, \ref{tab:sociodem_CI}, \ref{tab:sociodem_CT}, \ref{tab:sociodem_MR}, and \ref{tab:sociodem_SV} provide the corresponding sociodemographic breakdowns for each of the five locations.

Table~\ref{tab:sociodem_place} summarizes overall statistics for the thermal walks and how the data is distributed across the five locations, including the number of groups (thermal walks), participants, surveyed sites, and thermal surveys collected.

\subsection*{Working sessions with participants}

In each neighborhood and for each participant group, three participatory sessions were conducted over three consecutive weeks during May, June, or July 2024 as summarized in Figure \ref{fig:sessions}. The scheduling of these sessions was adapted to align with participants’ availability—for example, coordinating with school calendars or summer camp schedules. Each session lasted approximately 1.5 to 2 hours, with formats tailored to the specific characteristics and needs of each group. For instance, child-friendly formats such as games and quizzes were added to further engage younger participants. While delivery formats slightly varied to better suit each audience, the core content remained consistent across all sessions everywhere. The Heat Chronicles campaigns followed a community-based approach, with a dedicated team member—trained in community facilitation—ensuring an inclusive and welcoming environment throughout the process.

\subsubsection*{Session 1 (indoor and outdoor)}

The research team delivered a presentation introducing the concept of citizen science, outlining the rationale for the Heat Chronicles campaigns—including urban heat islands and heat inequity—and explaining their objectives along with the planned research steps.

\begin{figure}[H]
\centering
\includegraphics[width=1\textwidth]{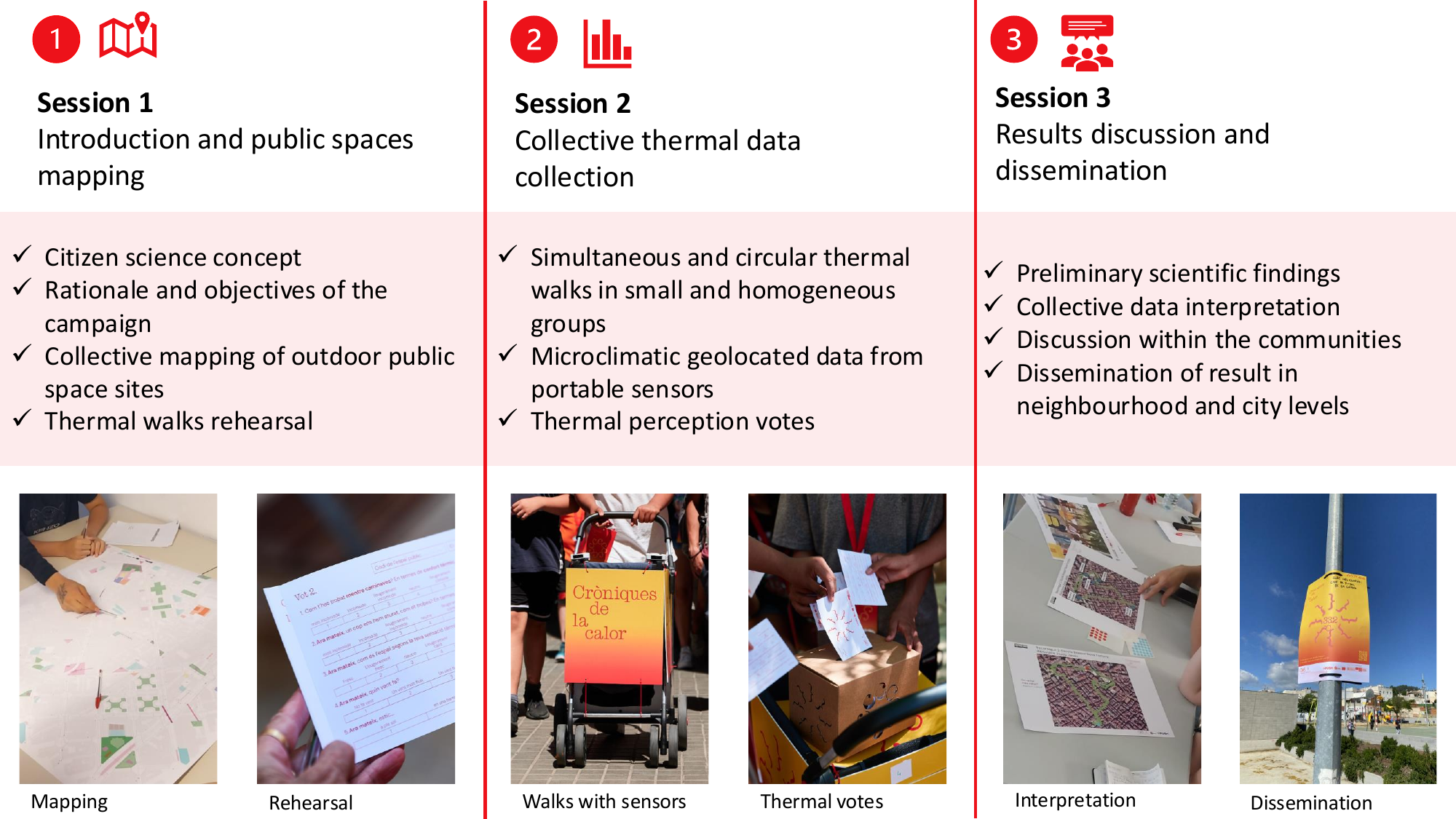}
\caption{{\bf Contents and activities of the three working sessions with participants.} (a) Introduction and public spaces mapping (Session 1); (b) Collective thermal data collection (Session 2); (c) Results discussion and dissemination (Session 3). Images illustrate the activities developed during these sessions.}
\label{fig:sessions}
\end{figure}

This was followed by two participatory and interactive activities as described in Figure \ref{fig:sessions}. In Activity 1, participants collaboratively mapped public spaces within their neighborhood, classifying and prioritizing them based on perceived relevance. The classification followed the categories provided by the research team. Large-format printed maps of each neighborhood (A0 size) were used during the activity (see Figure \ref{fig:sessions}). Participants identified and named relevant public space sites, marked them them on the map, and assigned them to one of the predefined categories using corresponding color codes. These categories were grouped into two overarching macro categories—{\it Basic Needs} and {\it Wellbeing and Free Time}—reflecting the different functions that public spaces serve in daily life as outlined in Table~\ref{tab:space_categories}. Each specific site was assigned a short code to support data labeling and visualization during a discussion phase. This classification scheme helped guide participants’ selection and prioritization of public spaces and ensured consistency in the thermal data collected across diverse urban environments. In Activity 2, the concepts of thermal sensation and thermal comfort were introduced. The session concluded with a collective rehearsal of the thermal walk data collection protocol, which is described in detail in a dedicated section below.

\subsubsection*{Session 2 (outdoor)} 
Thermal walks were conducted during this session, covering the prioritized sites identified by participants (see Figure \ref{fig:sessions}, and description in the following sections). The analysis focused exclusively on outdoor public areas and adjacent surroundings of indoor public sites, such as squares in front of primary health care centers, public libraries, or theaters. 

\begin{figure}[H]
\centering
\includegraphics[width=1\textwidth]{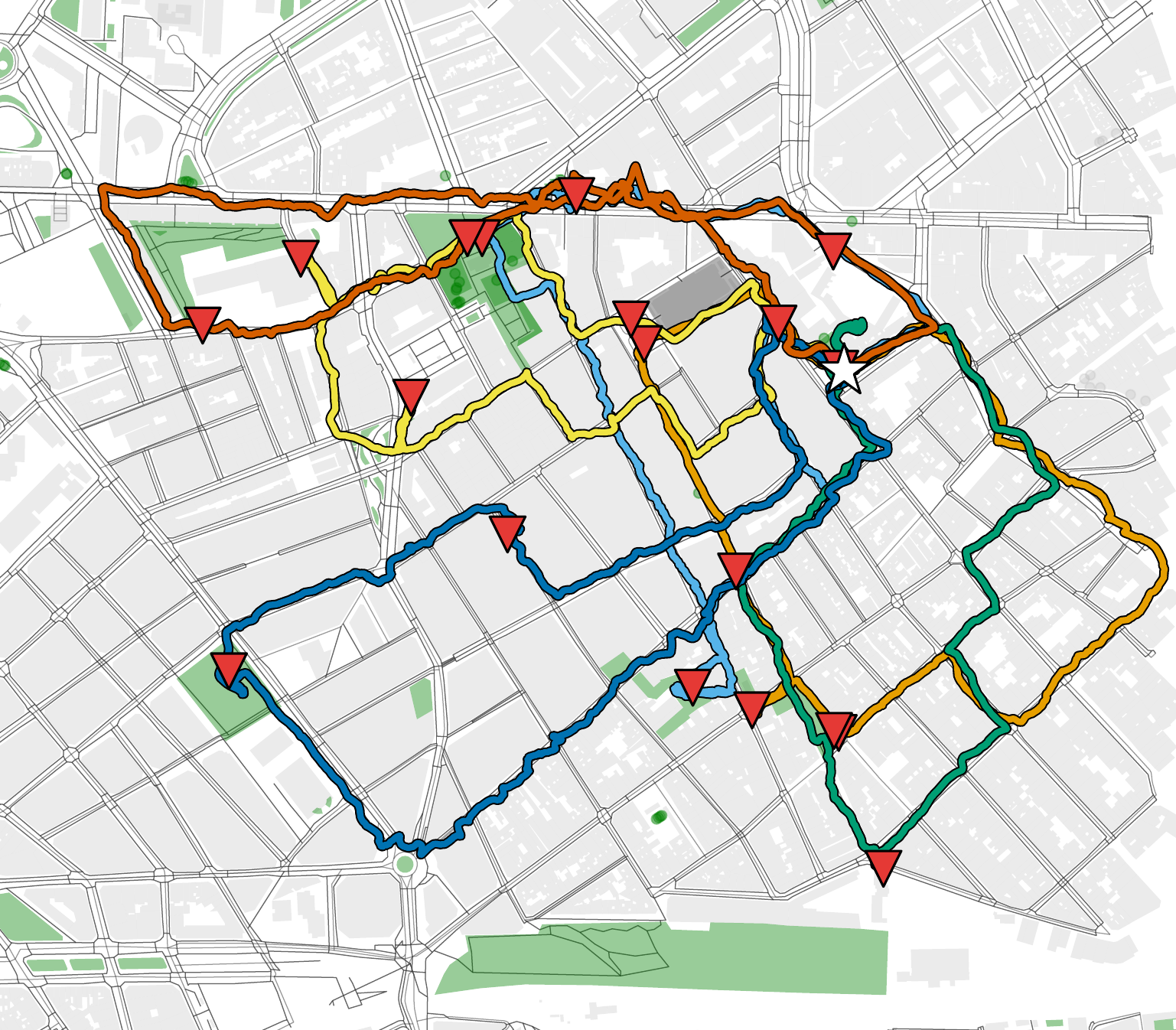}
\caption{{\bf Thermal walk routes in one location.} Simultaneous thermal walks conducted on May 30, 2024, in Collblanc-La Torrassa (CT, L'Hospitalet de Llobregat). Each participant group was assigned a different color to distinguish their route. Routes were designed to cover 31 outdoor public sites (red marks) prioritized by the community--including the starting/ending point (white star)--which were identified and labeled using predefined alphanumeric codes. A total of 67 participants were involved, divided into 6 groups corresponding to 6 different routes. Along these routes, thermal surveys collected a total of 959 self-reported entries (TSVs, TCVs and wTCVs).}
\label{fig:routes}
\end{figure}

Routes from each campaign were designed by the research team to include the identified sites agreed at the end of session 1. A member of the research team was then in charge to finetune all paths, maximizing the number of public sites visited and optimizing their feasibility, in terms of duration and total length of the trajectories. An example of simultaneous thermal walks for CT is presented in Figure \ref{fig:routes}. 

All participants convened at a designated starting point, and multiple groups conducted their walks simultaneously to collect both environmental measurements and perceptual data (see following sections for logistical and methodological details). Most thermal walks included between three and five stops, comprising the initial and final surveys at the starting point, as well as intermediate measurement sites. A few walks included more than five stops, resulting in additional measurement points beyond the start and end sites. Group sizes ranged from 6 to 12 participants. Approximately half of the walking trajectories spanned between $1.5$ km and $2.4$ km. The session was not longer than 2 hours despite spending much fewer time walking. Several minutes (see the data collection protocol below) were allocated for organizing the groups, reviewing the protocol, conducting thermal surveys—during which participants remained stationary—and closing the session.

\subsubsection*{Session 3 (indoor and outdoor)} 
Once the data collection was completed, the research team began by recapping the main characteristics of the thermal walks, including the itineraries, group organization, and participants' initial reflections. Preliminary scientific findings were also presented to stimulate discussion. In Activity 1, participants engaged in a collective interpretation of the data, drawing on their lived experience and local knowledge to contextualize the observed thermal patterns. This process included comparing sensor-based measurements with their own thermal perceptions and identifying relevant spatial features (e.g., shaded areas, open spaces, or material types). In some sessions, this was complemented by collaborative mapping or annotation exercises. Activity 2 (optional) involved preparing posters to communicate the results to the broader community. These posters summarized key findings and were displayed in relevant public spaces to raise awareness and foster dialogue on outdoor thermal comfort. Figure~\ref{fig:sessions} shows one of these posters.

In a later stage, some of the results contributed to neighborhood-level discussions on climate mitigation strategies during the neighborhood council in CI. In CT, participants featured in television news reports at both local and regional levels, as well as in a nationally broadcast documentary. Local organizations in CT are using the findings to advocate for concrete actions. Issues such as the lack of public drinking water fountains, shaded areas, and tree cover were clearly identified. Heat Chronicles was also invited by the municipality to participate in the Smart City Congress. In SV, the results were shared with both the municipal administration and the children’s council, fostering intergenerational dialogue. The citizen science campaigns also received local media attention, with several appearances on local radio. Specific demands were voiced to protect the football pitch frequently used by children from excessive heat and to improve accessibility to a nearby large green area. In SP, the CSO involved showed particular interest in sites highlighted during the thermal walks, as these locations serve as important gathering spaces for the community. The SP neighborhood is characterized by narrow streets and a significant lack of green areas. To support broader awareness and transparency, public reports summarizing the findings and key recommendations are being published and made accessible to the general public.

\subsection*{Thermal walks}
A thermal walk is defined here as a structured group walk along a pre-established route during which participants measure ambient temperature and systematically record their subjective perceptions of heat and thermal comfort \cite{Vasilikou_2020}. During the Heat Chronicles thermal walks, geolocated temperature, pressure, and humidity data were collected using MeteoTracker low-cost sensors, which have been shown to provide robust and sufficiently precise measurements in urban environments \cite{Barbano_2024}. Perceptual data were gathered through thermal surveys administered at stops located at predefined outdoor sites, following questionnaires similar to those employed in previous studies on outdoor thermal comfort \cite{Dzyuban_2022}. The survey instruments were developed in accordance with ISO 7730 \cite{ISO_7730} and ASHRAE Standard 55 \cite{ASHRAE_55}, including standardized scales for thermal sensation votes (TSVs), thermal comfort votes (TCVs) and thermal comfort votes while walking (wTCVs) .

\subsubsection*{Materials}
\begin{figure}[H]
\centering
\includegraphics[width=1\textwidth]{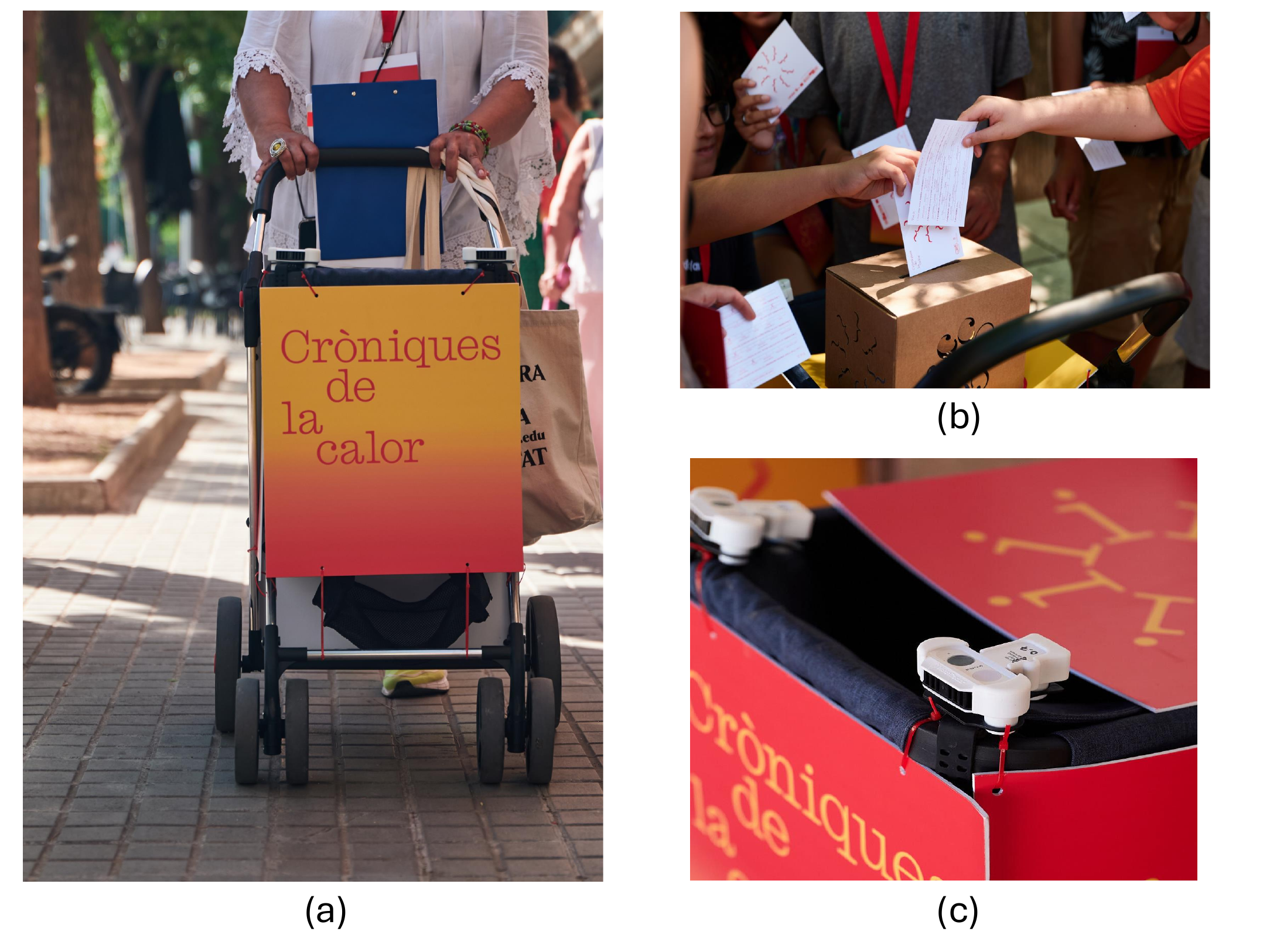}
\caption{{\bf Mobile equipment for the thermal walks.} (a) Modified shopping cart with custom panels and two sensors mounted consistently oriented to the direction of movement. (b) Ballot box and individual ballots used to capture participants’ subjective perceptions of heat. (c) Sensors connected via the MeteoTracker app to tablets inside the cart, enabling continuous recording of geolocated microclimatic data.}
\label{fig:croniques}
\end{figure}

We create our own materials to deploy the thermal walks. The experimental design was guided by three main principles: (i) collecting reliable scientific data, (ii) using lightweight and easy-to-transport materials, and (iii) making the experiment visible to the neighborhood while ensuring an enjoyable experience for participants. Based on these principles, the Heat Chronicles campaign graphic identity was developed, and a set of specially designed materials was produced (see Figure \ref{fig:croniques}).

The mobile equipment used for the thermal walks consisted of a commercially available shopping cart (Basket Tweed 4B Rolser) modified with specially designed panels for visibility and usability (see Figure \ref{fig:croniques}). Each cart was equipped with two MeteoTracker X sensors (https://meteotracker.com) mounted on top using the manufacturer’s bicycle-mounting system (Figure \ref{fig:croniques}c). This setup ensured that the sensors were consistently oriented in the direction of movement and positioned at a uniform height of 85 cm from the ground—chosen to approximate an average torso height across participants in the Heat Chronicles campaigns, which included both children (aged 9–11) and older adults. Two sensors were used per cart to increase measurement reliability, reduce fluctuations, and provide redundancy in case of sensor malfunction. The sensors were connected via the MeteoTracker mobile application to tablets (iPad, with the iOS operating system and 1GB of high-speed 4G connectivity) stored inside the cart, enabling continuous recording of geolocated microclimatic data (temperature, humidity, and pressure) throughout the walks.

To capture participants’ subjective perceptions of heat, a custom-designed ballot box was stored inside the cart (see Figure \ref{fig:croniques}b). Before starting a thermal walk, each participant received an envelope with a lanyard containing up to seven ballots and a pen. The ballots, whose graphic design is shown in Fig. \ref{fig:votes}, featured distinct drawings to facilitate easy identification. At each stop, participants completed a ballot reflecting their heat perception and deposited it into the ballot box. Each ballot included a randomly assigned participant ID, ensuring anonymity as the research team could not link IDs to individual participants.

\subsubsection*{Logistics and supervision}
The number of simultaneous thermal walks conducted in each location required careful logistical planning. One member of the team coordinated all organizational aspects, including preparing the materials and ensuring their timely shipment to the different sites.

The research team also arranged the composition of the participant groups, aiming for groups that were as homogeneous as possible in terms of age and of a manageable size, typically between 6 and 12 participants.

Altogether, up to 15 team members accompanied the different participant groups. Eight were members of the research team, while seven were students from the Faculty of Physics of the Universitat de Barcelona. All facilitators were trained to implement the scientific protocol rigorously and were provided with a printed guide detailing each step of the procedure. On the day of the walks, the team convened at the meeting point one hour in advance to prepare all materials. One person was specifically responsible for setting up the sensors and connecting the tablets. Before starting, the logistics coordinator conducted a briefing to ensure a shared understanding of the protocol and any relevant information about the neighborhood context.

In each participant group, a team member acted as the “group driver,” responsible for guiding the route and recording information in the driver’s sheet, translated into English in Table \ref{tab:driver_sheet}). For groups including children and youth under 18 years of age, a teacher or educator also accompanied the walk. Whenever possible, the drivers conducted a rehearsal walk in advance to familiarize themselves with the route and anticipate any challenges.

\subsubsection*{Thermal walk data collection protocol}
An equipped cart was placed in a shaded area at the meeting point where all simultaneous walks for a given location began and ended. A pair of “fixed sensors” remained in this site for the entire duration of the thermal walks in the location. These sensors were supervised by a team member to ensure a continuous and reliable record of temperature at a static reference point throughout the activity. All carts equipped with MeteoTracker sensors were positioned at the meeting points (starting and ending point of the walks), with sensors activated approximately 10 minutes before the walk began to ensure temperature alignment with ambient conditions.

At the meeting point, participants were divided into several groups. Each group was accompanied by one or two team members, including a designated group driver. The group driver introduced themselves and asked participants for their first names—although in most cases, they had already met during the first session. Participants were invited to take on different voluntary roles: guide (receiving a printed map with the route and stops), cart driver (responsible for pushing the equipment cart), and coordinator (ensuring all group members remained together). Roles could be rotated among participants during the walk.

\begin{figure}[H]
\centering
\includegraphics[width=1\textwidth]{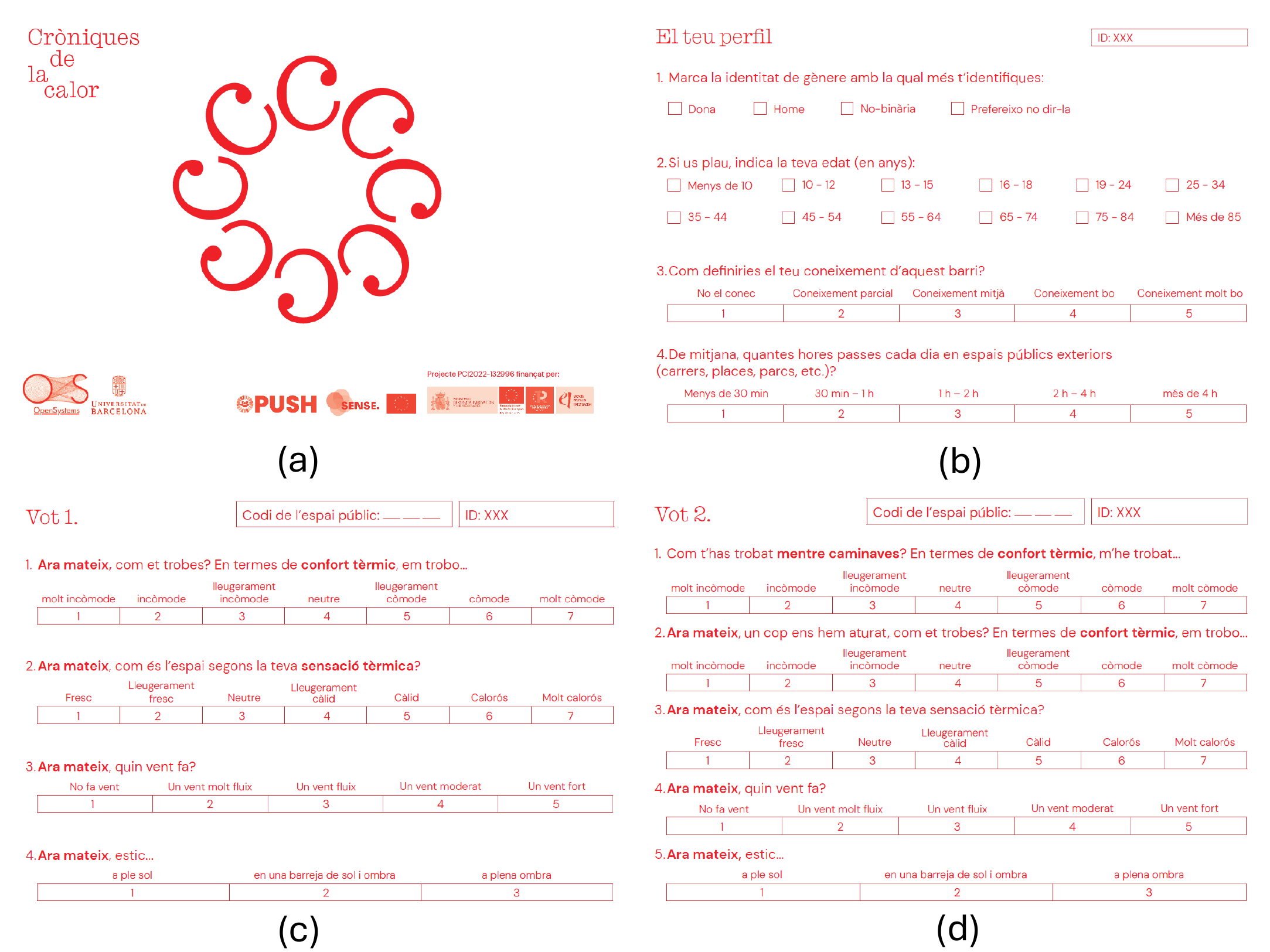}
\caption{
{\bf Paper ballots used for voting in the thermal surveys (in Catalan).}
(a) Example of the back of a ballot, displaying a unique design for easy identification. (b) Sociodemographic form completed by participants before the walk (see Table~\ref{tab:sociodem_english} for the English translation). (c) Survey on thermal comfort, thermal sensation, wind, and sun exposure from Ballot 1, completed before starting (see Table \ref{tab:vote1_english}). (d) From Ballot 2 onwards, the questions are the same as in (c) with an additional question on thermal comfort experienced while walking from the previous site (see Table \ref{tab:vote2_english}).}
\label{fig:votes}
\end{figure}

{\bf Sociodemographic survey:} Before departing, the group driver asked participants to complete the ballot corresponding to the sociodemographic survey (see Fig. \ref{fig:votes}b and Table \ref{tab:sociodem_english}). The driver also recorded general group information (such as the total number of participants and their approximate age range, see Table \ref{tab:driver_sheet}) on the driver’s sheet to allow cross-checking of the individual sociodemographic data. Completed ballots were deposited in the ballot box stored inside the cart.

{\bf First thermal sensation and comfort votes:} The group driver reminded participants of the distinction between thermal sensation and thermal comfort. Participants were then asked to complete (with no interaction with their peers) the first thermal perception survey, corresponding to Ballot 1 (see Fig. \ref{fig:votes}c and Table \ref{tab:vote1_english}). The driver announced the site code, which participants copied onto their ballots. This code consisted of a three-part sequence: the first number indicated the group number, the letter represented the date of the walk, and the final number identified the ballot (e.g, 1V1 for Ballot 1, and 1V2 for Ballot 2 in campaign V and cart 1). Once completed, all ballots were collected in the ballot box stored inside the cart.

{\bf Walk to the first and subsequent pre-identified sites:} The group proceeded together to the first designated site at a normal walking pace. The group driver and the participant assigned the role of guide indicated the path to follow. The same procedure applied to each subsequent transect between consecutive pre-identified sites.

{\bf Stop in a public outdoor pre-identified site:} The sites were pre-marked on the ground, geolocated using Google MyMaps, and shared with all drivers in advance. Once the group arrived at a pre-identified site, the group driver recorded the exact time and engaged participants in a brief conversation unrelated to heat. After 3–5 minutes of being stopped, the driver announced the corresponding site code, which participants wrote on their ballots. The waiting period was intended both to allow air temperature alignment and to help participants become fully aware of the thermal conditions at the site. The driver then asked participants to complete Ballot 2 (see Fig.\ref{fig:votes}d and Table \ref{tab:vote2_english}). Meanwhile, the driver filled in the driver’s sheet (see Table \ref{tab:driver_sheet}). All ballots were then collected in the ballot box. 

This procedure was repeated at each stop on the pre-identified sites. The final stop corresponded to the initial meeting point, as the routes were circular, with all groups converging after completing their thermal walks. At this final stop, participants completed the last round of votes.

The group drivers then invited participants to share feedback about the experience and reminded them of the date for the third session. The scientific team checked that all sensors had correctly recorded the journeys and that the data were successfully transferred to the MeteoTracker server. Finally, the sensors were disconnected, and the ballots from each box were placed in a separate envelope labeled with the group number and the name of the group driver. 

\subsection*{Data filtering and processing}
The data collected during the thermal walks can be organized into two distinct datasets. The first dataset comprises high-resolution, geolocated microclimatic measurements—trajectories—recorded by the MeteoTracker sensors throughout the walks. The second dataset contains the responses to the thermal surveys administered to participants during the walks. These surveys capture individuals’ perceptions of heat in outdoor public sites, their sociodemographic characteristics, and contextual information about the physical conditions of each survey site.

Regarding the MeteoTracker sensor data, we provide not only the raw trajectories but also processed trajectories prepared with specific filtering and cleaning procedures to facilitate further analysis. First, we cleaned the data by removing empty and irrelevant columns. Next, we applied linear interpolation to resample all records to a uniform temporal resolution of one measurement per second. To enhance data reliability, we then averaged the readings from the two sensors mounted on each cart. Finally, we appended additional derived columns, including temperature and humidity indices such as the humidex (HDX), which are described in detail in the following section.

With regard to the survey data, we digitized the responses from the thermal perception surveys completed by participants at the pre-selected sites. For each group and its corresponding trajectory, we compiled a dataset containing detailed information about each survey site, combining participants’ perceptions with microclimatic data extracted from the associated sensor readings. In addition to these group-level trajectory datasets, we generated two aggregated datasets: one that includes all survey stops across all thermal walks, and another that compiles all individual survey ballots (defined as a single participant’s response at a specific site). Finally, we created an additional dataset in which each record represents a unique participant ID, enabling participant-level analysis. The following sections provide a detailed explanation of the data processing steps and a comprehensive overview of the resulting datasets.

\begin{figure}[H]
\centering
\includegraphics[width=1\textwidth]{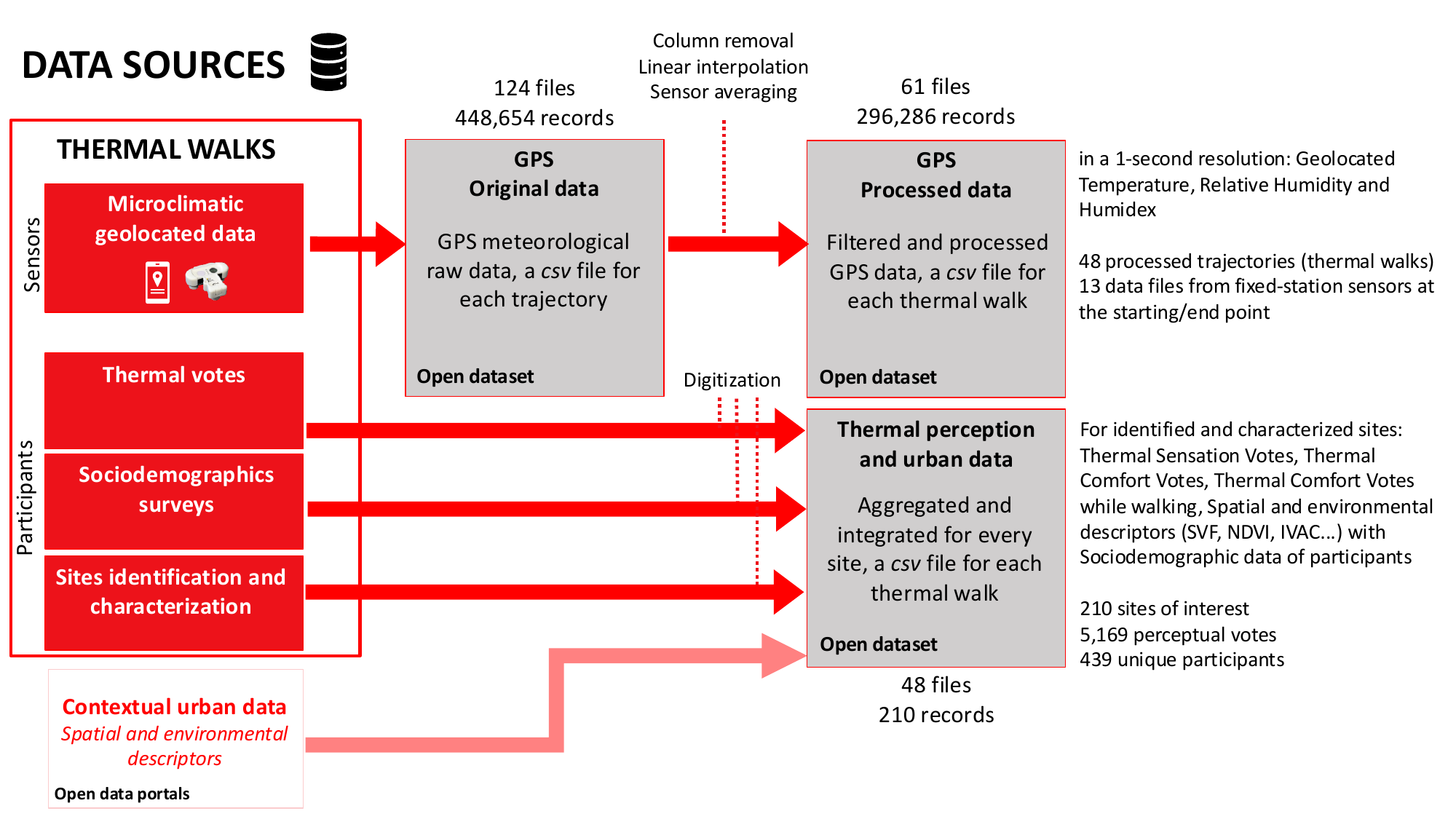}
\caption{{\bf Schematic diagram of data processing}. We provide both raw and processed individual trajectories, with geolocated microclimatic sensor data. Participants' heat perceptions at each site, along with their sociodemographic profiles, were digitized. This information was used to build a dataset for each thermal walk, containing data at each site. We enriched the dataset with urban contextual indicators for each site (e.g., NDVI, SVF, IVAC). Finally, we compiled three aggregated files: one with data for all 210 sites, another with all 1\,867 unique responses, and a third with the 439 complete response sets by participant ID.}
\label{fig:data_workflow}
\end{figure}

Figure~\ref{fig:data_workflow} illustrates the data flow, providing an overview of the processing and filtering procedures applied to the collected records.

\subsubsection*{Processing and filtering of sensor trajectories}\label{ss:trajectories_processing}
As mentioned in previous sections, two sensors were mounted on each cart to improve measurement accuracy and ensure reliable data collection in the event of sensor failure, connectivity loss, or poor data quality. The sensors recorded geolocated microclimatic data at a very high temporal resolution. Specifically, 59\% of the records were collected at one-second intervals and 41\% at two-second intervals. Only 0.03\% of records were separated by three or more seconds, likely due to transient GPS or Bluetooth connection issues, but these cases are statistically negligible.

\begin{figure}[H]
\centering
\includegraphics[ width=1\textwidth]{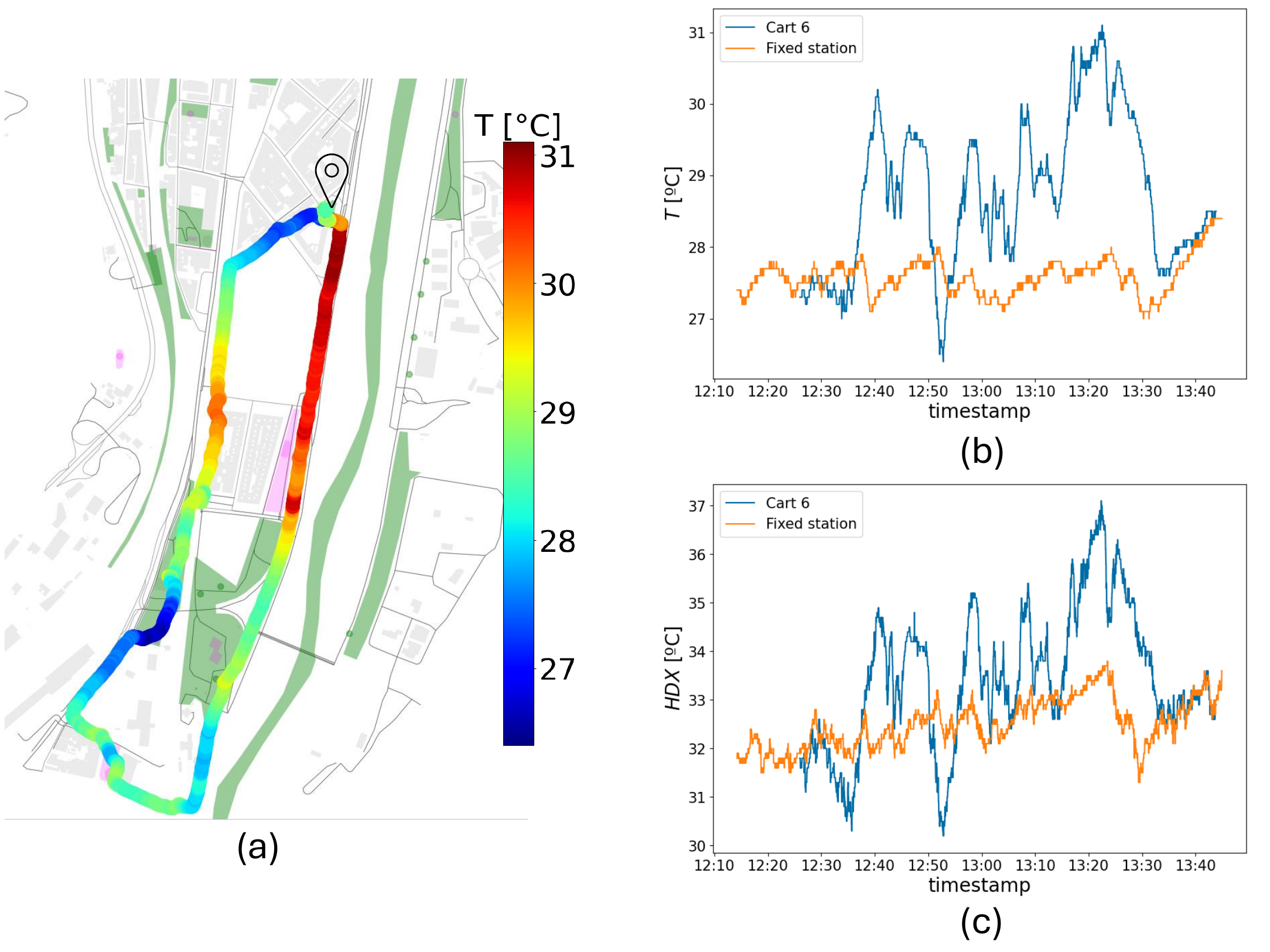}
\caption{\label{fig:data_visualization} \textbf{Example visualization of a thermal walk.} Thermal walk conducted by a group in MR, using data collected with the MeteoTracker sensor. (a) Heat map showing air temperature along the walking route, which started and ended at La Ribera Institute. (b) Temporal evolution of air temperature during the walk. (c) Temporal evolution of humidex during the walk.}
\end{figure}

Across all raw trajectories from all sensors, the dataset comprises 124 unique files and a total of 448\,654 records. In 6 out of the 61 trajectories, one of the two sensors mounted on a cart failed for various reasons. Figure \ref{fig:data_visualization} illustrates a sample thermal walk, showing a heatmap of geolocated temperature readings alongside the time series of temperature and humidex (HDX) recorded during the walk.

{\bf Step 1: Remove empty and irrelevant columns.} Each individual dataset initially contained 28 columns. Most of these columns were empty or corresponded to functionalities not supported by the MeteoTracker sensor model we used. Therefore, we retained only the 10 relevant columns (see Table~\ref{tab:columns_raw_trajectories}): Time, Lat, Lon, Temp [$^\circ\text{C}$], Hum [$\%$], Alt [m], Press [mbar], DP [$^\circ\text{C}$], HDX [$^\circ\text{C}$], and Speed [km/h]. Table~\ref{tab:example_trajectory_file} shows an example of a raw trajectory file.

\begin{figure}[H]
\centering
\includegraphics[width=1\textwidth]{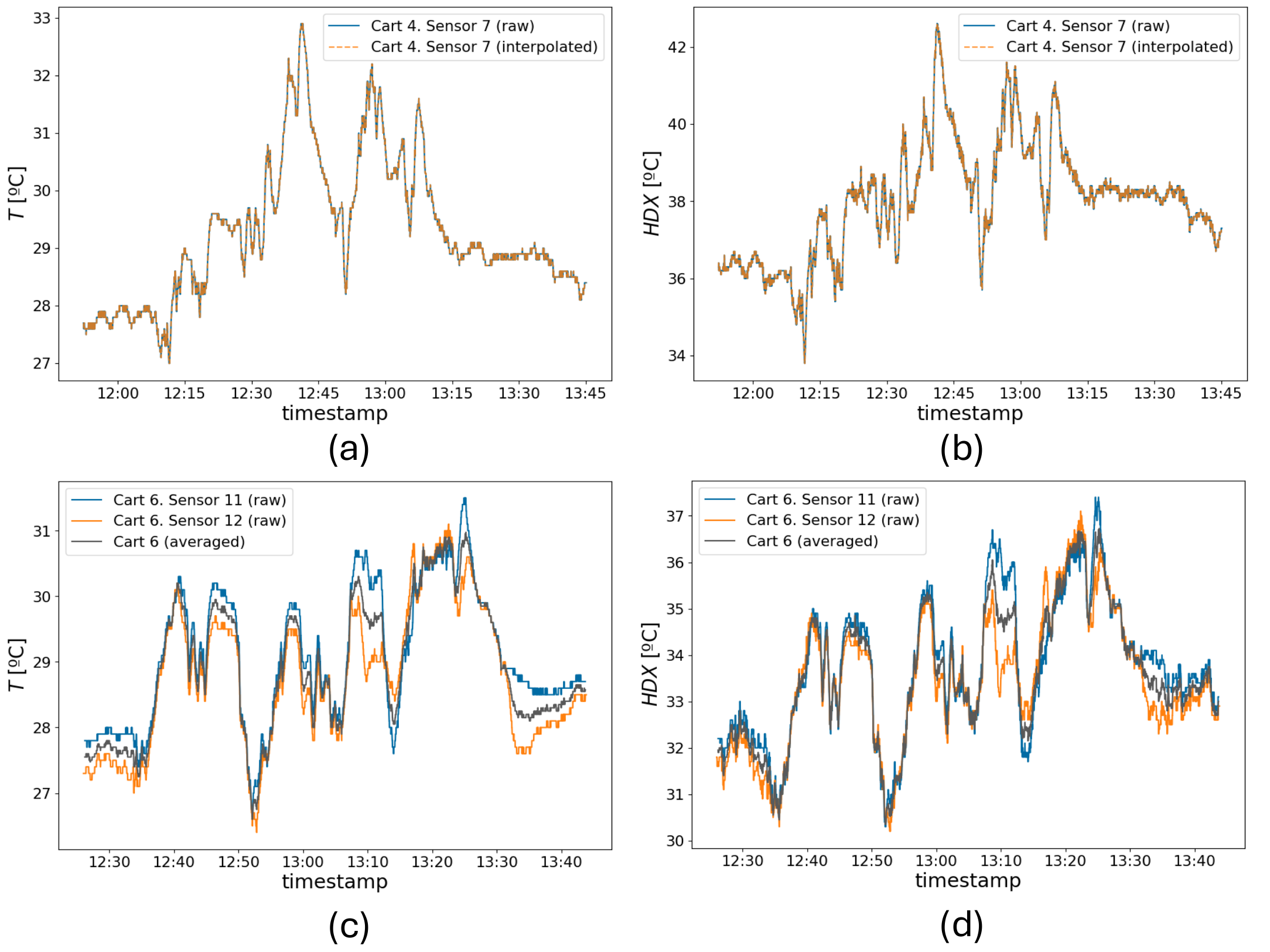}
\caption{{\bf Temporal evolution of temperature and humidex.} (a) Temporal evolution of temperature during a thermal walk in CT, on July 9, 2024. The blue line represents the raw temperature data, while the orange line shows the result after linear interpolation. (b) Same as (a), but showing the humidex (HDX). (c) Temporal evolution of temperature during a thermal walk in MR, on June 7, 2024. Blue and orange lines correspond to the original temperature data from the two sensors installed on the same cart. The black line shows the processed temperature, obtained by averaging both sensors' data at each timestamp. (d) Same as (c), but showing the humidex (HDX).}
\label{fig:technical_val1}
\end{figure}

{\bf Step 2: Perform linear interpolation.} For many statistical analyses, it is preferable to work with uniformly spaced records. This also facilitates averaging measurements from the two sensors mounted on each cart, since the variables must be aligned to the same timestamps. Therefore, we applied a linear interpolation procedure to fill all gaps where consecutive records were separated by more than one second. This processing ensured that all records were evenly spaced in time (at one-second intervals). After interpolation, the total number of records increased from 448\,654 to 637\,619 (a 40\% increase). Although the number of records grew substantially, the overall data distributions remained unaffected (see {\tt Technical Validation} section, Figure~\ref{fig:technical_val1}a and Figure~\ref{fig:technical_val1}b).

{\bf Step 3: Average the two sensors from the same cart.} To improve data reliability, we averaged the readings from the two sensors mounted on each cart at every timestamp (1-second intervals). This averaging was performed over the overlapping time interval between the two sensors—defined as the period from the latest starting time to the earliest ending time where both sensors were actively recording. In cases where a cart collected data from only one sensor, we retained the available single-sensor data without averaging. As a result, the number of individual trajectories was reduced from 124 to 61, with a total of 296\,286 records in the processed dataset. The \texttt{Technical Validation} section and Figure~\ref{fig:technical_val1}c and Figure~\ref{fig:technical_val1}d illustrate how this averaging procedure reduces noise in both temperature and humidex measurements.

{\bf Last step: Shift temperatures.} We corrected the temperature data to mitigate the effect of natural temperature variation over the course of the day. During a given trajectory, ambient temperature may gradually increase or decrease due to the diurnal cycle. As a result, a sensor might capture this temporal trend, potentially leading to misleading interpretations when comparing temperatures between two sites that were recorded at different times along the same trajectory. We thus shifted the temperature at each timestamp $t_i$ by subtracting the temperature recorded by the fixed-station sensor located at the starting (and ending) point of the trajectory:
\begin{equation}
T_s(t_i) = T(t_i) - T_{fixed}(t_i).
\label{eq:rescaledT}
\end{equation}
This subtraction removes broad temporal trends from spatial temperature differences, which typically results in small temperature changes (on the order of a few decimal degrees), either positive or negative. A second optional adjustment adds back the average temperature recorded by the fixed-station over the entire duration of the trajectory, shifting the data back to a range more representative of actual atmospheric conditions while preserving relative spatial differences:
\begin{equation}
T_{s'}(t_i) = T(t_i) - T_{fixed}(t_i) + \langle T_{fixed}(t_i) \rangle.
\label{eq:rescaledT2}
\end{equation}
We applied the same procedure to the humidex, thereby adding four additional columns to each trajectory dataset (two for temperatures and two for humidex).

\textbf{Processing and filtering fixed-station sensor data.} As described above, a cart equipped with two sensors was placed at the starting point of each thermal walk, which also served as the ending point (i.e., start and end coincide). For these fixed-station datasets, we repeated steps 1, 2, and 3 described above. As a result, the fixed-station datasets contain 10 columns (Table \ref{tab:columns_raw_trajectories}).

After processing the sensor data from all trajectories, we obtained a total of 61 cleaned and unique datasets. Of these, 48 correspond to geolocated trajectories (thermal walks), each containing 14 columns: the 10 microclimatic variables recorded by the sensors plus 4 additional variables derived through data processing (see Last step). The remaining 13 datasets correspond to the fixed-station sensors, each containing the same 10 columns.

\subsubsection*{Thermal surveys responses}\label{ss:stops_processing}
We digitized all individual responses from the thermal perception surveys, along with the driver sheets corresponding to each thermal walk. To ensure data accuracy, multiple verification steps were carried out to identify and correct any potential human errors introduced during the manual data entry process.

In the first stage, for each thermal walk, we created a dedicated dataset containing the thermal survey responses of the group. Each row corresponds to a stop at a site where a survey was conducted, and each stop is characterized by 80 columns. This process yielded 48 individual datasets—one for each group—and a consolidated dataset aggregating all responses, totaling 210 surveys across all geolocated outdoor public sites.

{\bf The sites.} For each stop of each trajectory, the first 13 columns include the site code (e.g., 1V4), the name of the outdoor public site, the primary and secondary categories of the site (e.g., health, education), the city or neighborhood, the driver’s anonymized identifier, the date, the arrival time, and the departure time (both in HH:MM format). They also include the precise GPS coordinates (latitude and longitude) of the stop, obtained using Google MyMaps. Additionally, we derived GPS coordinates from the MeteoTracker sensor data by averaging the latitude and longitude recorded during the stop’s time window (from arrival to departure) in the corresponding group trajectory dataset.

{\bf The survey.} We added 51 additional columns. We first captured information recorded by the driver at the beginning of the walk and at each stop (see Table \ref{tab:driver_sheet}). These columns include the number of participants in the group, participants’ ages, how they were dressed, the presence of bad smells at the survey site, the noise level, the number of people within a 10-meter radius, and the total number of ballots collected at each stop (this total is later computed by summing the ballors). The next set of columns represents the counts of each possible response option for every question in the thermal survey (see Fig. \ref{fig:votes} and Tabs. \ref{tab:vote1_english} and \ref{tab:vote2_english}). Specifically, we created separate columns for each of the seven possible responses for thermal comfort (TCV), seven for thermal sensation (TSV), and so on. The first stop was conducted at the initial site before walking began. For this reason, the question about thermal comfort while walking does not appear on the ballot 1; this field is recorded as NaN in the dataset. Finally, for each question in the thermal survey, we computed and added three common statistical metrics as additional columns: the mode, the median, and the interquartile range (IQR). The mode is simply the most frequent response; if multiple responses tied for the highest count, all modes are stored as a list. The median identifies the response corresponding to the 50th percentile when all votes are sorted (for example, for thermal comfort, from “very uncomfortable” to “very comfortable”). If there were 12 votes, the 6th vote determines the median; in the case of an odd number of votes (e.g., 11), we considered the responses at ranks 5 and 6 and selected the most frequent as the median. The interquartile range (IQR) captures the variability by recording the 25th and 75th percentile responses. For example, with 11 votes, the 25th percentile corresponds to the 3rd vote (2.75 rounded up), and the 75th percentile to the 8th vote (8.25 rounded down). Both values are saved together as a list.

\textbf{Temperature and Humidex.} We included information on the ambient temperature and humidex at each site where the group stopped. To obtain these values, we used data from the sensor trajectories and computed the average for each variable over the time window defined by the group’s arrival and departure times at the site. The same procedure was applied to the shifted temperature values (cf. Eqs. (\ref{eq:rescaledT}) and (\ref{eq:rescaledT2})). This process added a total of six columns: three for temperature (raw, shifted, and shifted with offset) and three for humidex.

\textbf{Contextual urban information.} The last 10 columns describe the contextual and urban characteristics of the outdoor urban public sites corresponding to each stop. These columns include: the type of surface; the walking distance (in meters) to the nearest public fountain and to the nearest public drinking fountain (data source: \url{https://opendata-ajuntament.barcelona.cat/data/en/dataset/fonts}); the distance to the closest urban green space (data source: \url{https://opendata-ajuntament.barcelona.cat/data/ca/dataset/arbrat-parcs}); and the Sky View Factor (SVF, data source: \url{https://ide.amb.cat/Visor/?locale=en }), which measures the percentage of visible sky (a higher SVF indicates more open areas) \cite{Miao_2020}. We also include the Normalized Difference Vegetation Index (NDVI, data source: \url{https://www.icgc.cat/en/Data-and-products/Imatge/NDVI }), which estimates the amount, quality, and growth of vegetation based on the normalized difference between reflected infrared and red light \cite{Gascon_2016}. Additionally, the index of vulnerability to climate change (IVAC, data source: \url{https://www.amb.cat/web/area-metropolitana/dades-espacials/detall/-/serveidigital/index-de-vulnerabilitat-al-canvi-climatic--ivac-/13903812/11692}) \cite{Garcia_2022} and the walking distance to the nearest climate shelter are provided (data source: \url{https://opendata-ajuntament.barcelona.cat/data/en/dataset/xarxa-refugis-climatics}). Finally, two more columns report the mean SVF and mean NDVI values calculated over a 1.5-meter radius area. 

In the second stage, we compiled a comprehensive dataset containing all individual surveys collected at each stop. Each row represents a unique participant’s response to the thermal survey conducted at a specific site (see Figure \ref{fig:votes}). The dataset includes a total of 1\,867 surveys and 21 columns, capturing: the participant ID, sociodemographic details (gender, age, time spent in the public space, and neighbourhood familiarity), the location (city or neighbourhood), the site code (stop), and responses to each survey question (thermal comfort, thermal comfort while walking, thermal sensation, wind perception, and sun exposure). Additionally, the dataset reports the average temperature and humidex recorded for each stop, along with the corresponding shifted values (cf. Eqs. (\ref{eq:rescaledT}) and (\ref{eq:rescaledT2})).

Finally, we created a unique dataset with the same structure as the previous one (which contains all individual responses), but with each record corresponding to a unique participant ID. This dataset includes a total of 439 participants. Unlike the previous dataset, where each row represents a single record at a specific site, here the thermal survey responses are aggregated per participant. Each thermal-related variable is represented as a list of the participant's responses across all sites. For example, the thermal comfort field for a given participant might appear as: \textit{[Neutral, Very uncomfortable, Neutral, Comfortable, Slightly uncomfortable]}. Since temperature and humidex values are tied to specific sites and not individual participants, they are omitted from this dataset, resulting in 15 columns instead of 21.

\subsection*{Code availability}
The \url{https://github.com/ferranlarroyaub/Croniques_de_la_Calor_2024} repository contains the Python code and scripts for processing the input data and for replicating the data processing and the figures in this manuscript. The $3.8$ Python version is used to build the code with the main libraries networkx and osmnx to plot the trajectories on OpenStreet maps, pandas and numpy to process, clean and analyze the data in Data-Frame format and performing the basic statistical calculations, and matplotlib for plotting figures. The Python code is built in different Jupyter notebook files which contain the detailed description of the study and the code documentation.

\section*{Data Records}
The data repository \cite{Croniquesdata2025} contains both raw and processed datasets, as described in the \texttt{Data filtering and processing} section. The repository is organized into four main folders. The \texttt{original\_trajectories} folder includes the individual trajectories for each group, with geolocated microclimatic data collected using the MeteoTracker sensors during the thermal walks. The {\tt processed\_ trajectories} folder contains the same trajectories after undergoing the data processing procedures. The \texttt{processed\_surveys} folder stores all survey responses associated with each group (trajectory). Finally, the \texttt{aggregated\_ surveys} folder contains three files summarizing the aggregated survey data.

\subsubsection*{Raw trajectories}
All the original trajectories for each participating group are stored in the folder named \texttt{original\_trajectories}. The dataset comprises 124 unique trajectories, with a total of 448\,654 records of microclimatic data captured by the MeteoTracker sensors. Each file is provided in \texttt{csv} format and can be identified by the site where the data were collected (see Table \ref{tab:city_neighborhood}), the date, the assigned cart (group) number, and the sensor number. For example, the file \texttt{MR\_7june2024\_cart\_1\_sensor\_2.csv} corresponds to the trajectory recorded by sensor 2 of cart (group) number 1, in \textit{Montcada i Reixac (MR)}, on June 7, 2024. Another example is \texttt{CT\_9july2024\_fixed\_cart\_sensor\_11.csv} which corresponds to the data recorded by sensor 11 of the fixed-station, in \textit{Collblanc-La Torrassa (L'Hospitalet de Llobregat, CT)}, on July 9, 2024.

As detailed in the experimental protocol and data processing sections, the carts were equipped with two sensors to enhance measurement reliability. Upon examination of the file names, it is easy to identify the two sensors assigned to the same cart. For example, the files \texttt{ SV\_29may2024\_cart\_2\_sensor\_3.csv} and \texttt{SV\_29may2024\_cart\_2\allowbreak\_sensor\_4.csv} correspond to the same trajectory (cart number 2), recorded using sensors 3 and 4, respectively, for a group from \textit{Sant Vicenç dels Horts (SP)}, on May 29, 2024. Table \ref{tab:example_trajectory_file} shows an example of a \texttt{csv} file display of an individual trajectory. 

\subsubsection*{Filtered and processed trajectories}
The \texttt{processed\_trajectories} folder in the repository contains the final set of 61 unique trajectories, obtained after the filtering and processing procedure described above. Out of the 61 files, 48 correspond to the walking trajectories (with 14 columns each), and 13 to the fixed-stations (with 10 columns each), both described in \texttt{Data filtering and processing} section. The total number of records is 296\,286. As with the raw trajectories, each file is in \texttt{csv} format and can be identified by the location where the data were collected, the date and the assigned cart (group) number. The difference is that, in this case, the sensor number is not identified, as the file represents the result of averaging the data from the two sensors on the same cart. For example, the original trajectories \texttt{SV\_29may2024\_cart\_2\_sensor\_3.csv} and \texttt{SV\_29may2024\_cart\_2\allowbreak\_sensor\_4.csv} are combined into a processed dataset called \texttt{SV\_29may2024\allowbreak\_cart\_2.csv}.

\subsubsection*{Surveys data}
The \texttt{processed\_surveys} folder contains 48 \texttt{csv} files, each corresponding to a thermal survey conducted by a group during their thermal walk. Each record in these files represents a unique stop, where all participants responded to questions about their heat perception at a specific site. Each dataset includes 80 columns, covering: (i) information about the stop (e.g., site code, category of the public site, date, and time), (ii) details of the thermal survey (e.g., the number of responses for each possible answer to each question), and (iii) contextual and urban characteristics of the site. The average temperature and humidex values at each stop—calculated from the group’s sensor trajectory—are also included. Each \texttt{csv} file is identified by the location where the data were collected, the date, and the assigned cart (group) number, adding the suffix \texttt{survey}. For example, the file \texttt{CI\_4july2024\_survey\_cart\_1.csv} corresponds to the thermal survey conducted by group (cart) number 1, in \textit{El Congr\'es i els Indians (Barcelona, CI)}, on July 4, 2024.

\subsubsection*{Aggregated surveys data}
We provide three \texttt{csv} files with aggregated survey data in the \texttt{aggregated\allowbreak\_surveys} folder. First, the file \texttt{all\_surveys(stops).csv} is a direct aggregation of all 48 files from the \texttt{processed\_surveys} folder. It contains data from all 210 sites where surveys were conducted, with the same 80 columns as the individual files. The file \texttt{all\_surveys(votes).csv} contains all individual responses. Each record corresponds to a unique participant’s response at a specific site. As described in the \texttt{Data filtering and processing} section, this dataset includes 21 columns covering participants’ sociodemographic information (gender, age, time spent in the public space, and neighborhood knowledge) along with their answers to each survey question. Additionally, the average temperature and humidex values for each site are included. The last file, \texttt{all\_surveys(ID).csv}, contains the sociodemographic profile and thermal survey responses for each unique participant ID. This dataset has 439 unique participants, with columns similar to those in \texttt{all\_surveys(votes).csv} but excluding temperature and humidex data, resulting in 15 columns total.

\section*{Technical Validation}\label{s:technical_validation}

\subsection*{Temperature and relative humidity sensor data}
\begin{figure}[H]
\centering
\includegraphics[width=0.83\textwidth]{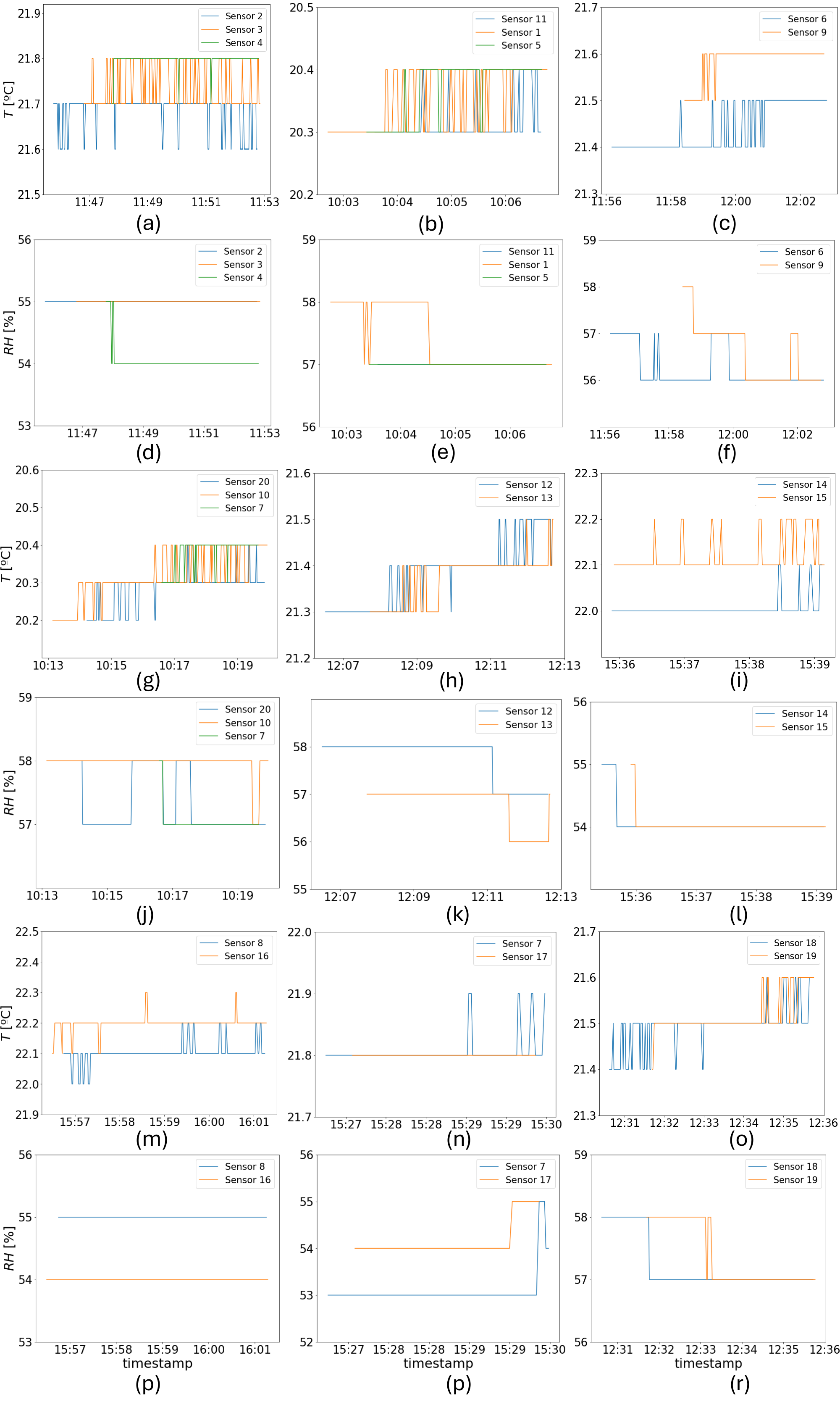}
\caption{{\bf Indoor sensor data validation.} Temperature and humidity readings across different combinations of sensors. All agree within the calibration precision specified by the MeteoTracker documentation ($\pm 0.2\,^\circ \mathrm{C}$ for temperature and $\pm 3\%$ for relative humidity).}
\label{fig:technical_val_calibration3}
\end{figure}

\begin{figure}[h]
\centering
\includegraphics[ width=1\textwidth]{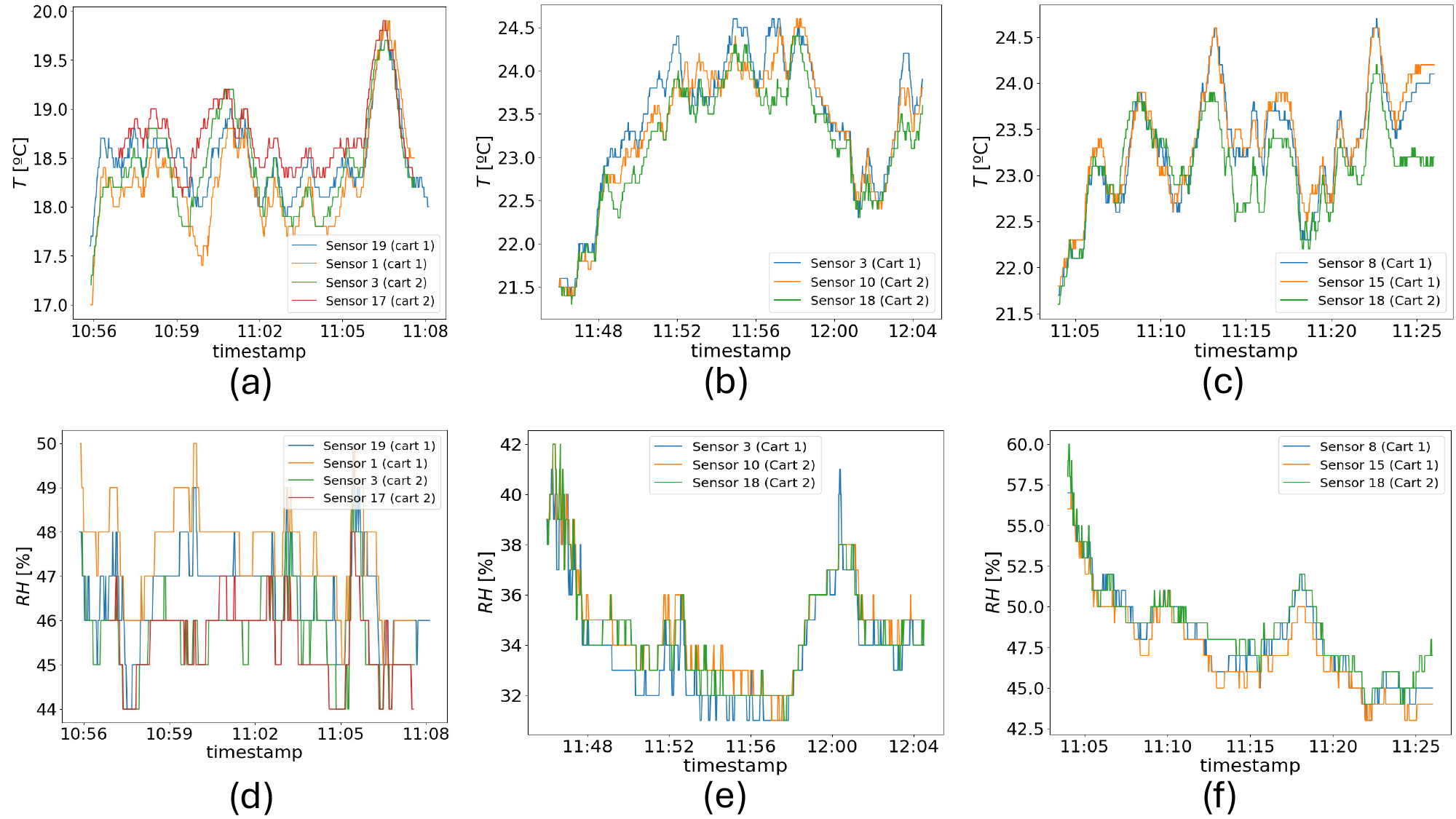}
\caption{{\bf Outdoor sensor data validation while walking.} Top panels (a-c) correspond to the time evolution of temperature for different combinations of sensors. Bottom panels (d-f) display the corresponding relative humidity for the same sensor sets. Temperature and humidity readings agree across sensors within the calibration precision specified by the MeteoTracker documentation ($\pm 0.2\,^\circ \mathrm{C}$ and $\pm 0.4 \,^\circ \mathrm{C}$ under solar radiation with speed $>7$\,km/h for temperature and $\pm 3\%$ for relative humidity).}
\label{fig:technical_val_calibration4}
\end{figure}

Prior to the thermal walks, the twenty MeteoTracker sensors were tested to verify their proper functioning and to ensure consistent microclimatic measurements when multiple sensors operated concurrently. To this end, a series of tests employing different sensor combinations were conducted, during which average temperature ($T$) and relative humidity ($RH$) values, as well as their temporal variations, were analyzed both indoors (within an isolated office environment) and outdoors (while traversing the streets of Barcelona using the carts). Results from the indoor tests, presented in Figure \ref{fig:technical_val_calibration3} and summarized in Table \ref{tab:calibration_averages}, demonstrate consistent sensor performance, with measurements agreeing within the manufacturer’s specified accuracy of $\pm 0.2\,^{\circ}$C for temperature ($\pm 0.4\,^{\circ}$C under solar radiation with speed $>7$\,km/h) and $\pm 3\%$ for relative humidity (\url{https://meteotracker.com}) and reports from a group of researchers \cite{Barbano_2024,Loglisci_2024}. The outdoor tests, conducted under field conditions, were inherently more susceptible to environmental influences such as solar radiation and wind exposure. Although not all sensors participated in these outdoor trials, various sensor combinations produced satisfactory and consistent results, as evidenced by the data presented in Table \ref{tab:calibration_averages} and Figure \ref{fig:technical_val_calibration4}, which depict average temperature and relative humidity values alongside their temporal dynamics. Minor discrepancies observed are attributed to environmental factors, including differential exposure to wind—such as when a cart trails another—and variations in sunlight exposure experienced by individual sensors while navigating urban pathways.

\subsection*{Geolocated data}
\begin{figure}[h]
\centering
\includegraphics[width=1\textwidth]{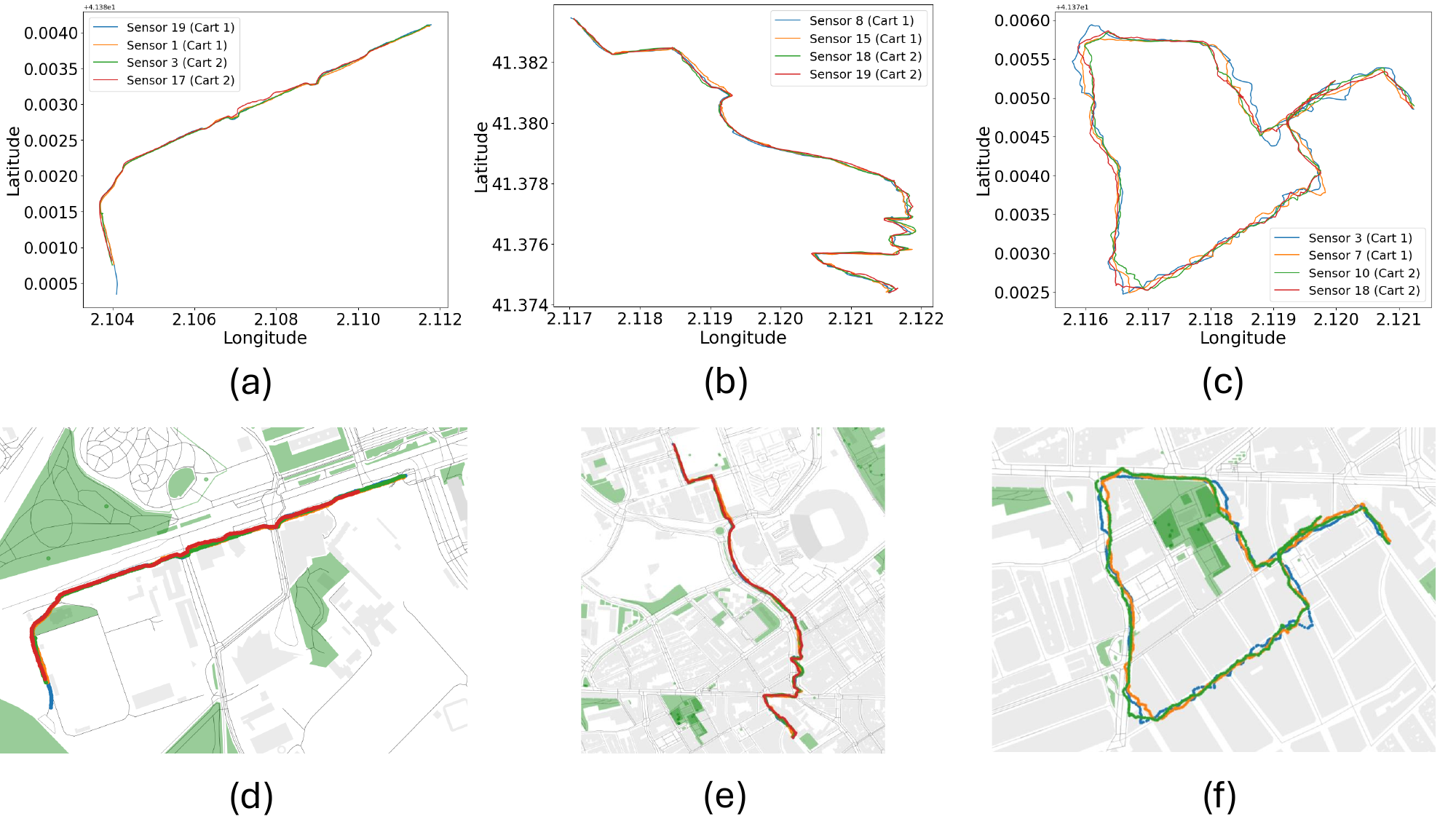}
\caption{{\bf Outdoor geolocation data validation.} Top panels (a–c) show GPS trajectories (longitude vs. latitude) for three separate tests, each involving different combinations of sensors. Bottom panels (d–f) display the corresponding geolocations projected onto maps using the OpenStreetMap layout, illustrating the same sensor sets as in the top panels.}
\label{fig:technical_val_gps}
\end{figure}

The geolocation of each sensor was obtained via Bluetooth using an iPad tablet device, as outlined in the Methods. During the outdoor tests, we verified that the GPS coordinates recorded by the sensors were generally accurate. Minor discrepancies between sensor locations can be attributed to standard GPS precision limitations, as well as occasional spatial and natural separation of a few meters between carts during movement. Figure \ref{fig:technical_val_gps} presents the GPS trajectories (latitude and longitude) from three separate walking tests, during which four sensors—mounted on two carts—collected data simultaneously. The recorded trajectories demonstrate a high degree of alignment in all tests, indicating consistent geolocation performance. However, the third test (Figures \ref{fig:technical_val_gps}c and \ref{fig:technical_val_gps}f) exhibits slightly noisier GPS signals. This increased variability can probably be attributed to the urban environment, characterized by narrow streets and tall buildings, which are known to degrade GPS satellite accuracy. Furthermore, the high population density of the area, combined with limited sidewalk space, frequently caused the two carts to become spatially separated by several meters.

\subsection*{Linear interpolation}
In the \texttt{Data filtering and processing} section, we detailed the procedures implemented to clean and preprocess the geolocated microclimatic data collected via MeteoTracker sensors during thermal walks. A fundamental step in the processing workflow was the application of linear interpolation to standardize the temporal resolution of the dataset, ensuring that all records were uniformly spaced at 1-second intervals. This interpolation was applied not only to temperature and humidex values but also to GPS coordinates, thereby generating interpolated positions along the straight line connecting two recorded locations separated by more than one second. Given the high spatial and temporal resolution of the raw data, this interpolation does not compromise the spatial or statistical integrity of the trajectories. Furthermore, the use of latitude and longitude expressed in degrees remains valid at this resolution, as the spatial differences between consecutive points are sufficiently small to approximate the coordinate system as linear for interpolation purposes.

\begin{figure}[H]
\centering
\includegraphics[width=1\textwidth]{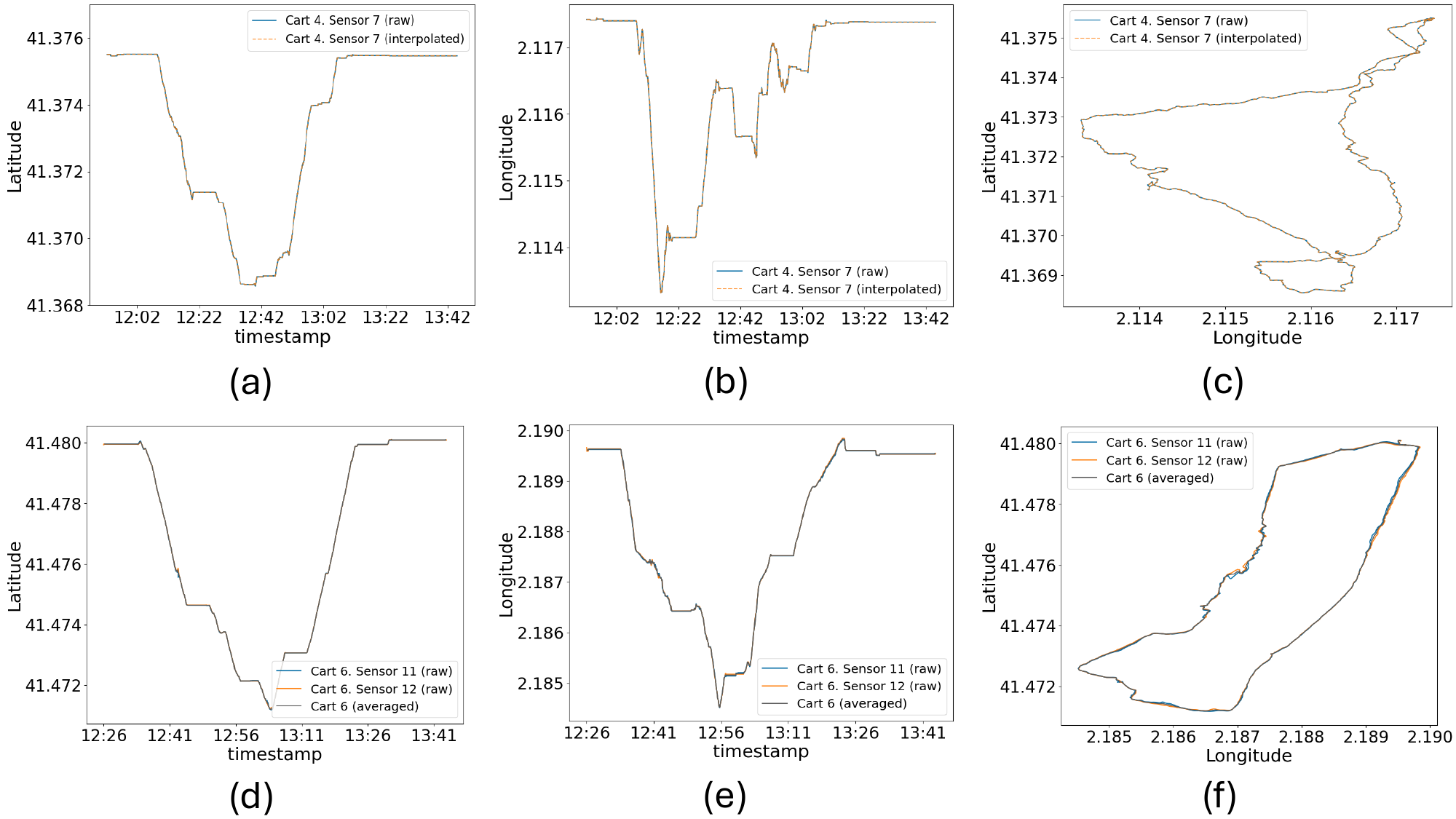}
\caption{{\bf Projected trajectory and temporal evolution.}
(a) Temporal evolution of the latitude coordinate during a thermal walk in CT. The blue line represents the raw GPS data, while the orange dashed line corresponds to the linearly interpolated values. (b) Same as (a), but for the longitude coordinate. (c) GPS trajectory (longitude vs. latitude) for the same trajectory, comparing interpolated and non-interpolated data. Bottom panels (d–f) present the same information, but comparing the raw GPS data from two sensors mounted on the same cart (blue and orange lines), along with the processed GPS trajectory obtained by averaging the two sensors' coordinates at each timestamp (black line).}
\label{fig:technical_val_gps2}
\end{figure}

Figures \ref{fig:technical_val1}a and \ref{fig:technical_val1}b illustrate the temporal evolution of temperature and humidex (HDX), respectively. Figure \ref{fig:technical_val_gps2}c displays the projected trajectory (latitude versus longitude), alongside the temporal evolution of both spatial coordinates (Figures \ref{fig:technical_val_gps2}a and \ref{fig:technical_val_gps2}b). A visual comparison between interpolated and non-interpolated data confirms that the overall trajectory patterns remain effectively unchanged. The average temperature and humidex values are identical across both datasets, with $\langle T \rangle = 29.28 \pm 0.02\,^\circ\mathrm{C}$ and $\langle HDX \rangle = 38.02 \pm 0.02\,^\circ\mathrm{C}$. Similarly, the total distance traveled along the trajectory—calculated as the sum of distances between consecutive GPS points—remains consistent at $D = 2\,481.94$ meters.

\subsection*{Average two sensors from the same cart}
An additional data processing step implemented to improve the accuracy of the collected measurements involved averaging the readings from the two sensors mounted on the same cart—applicable to nearly all thermal walks, as most were conducted using two sensors per cart. Despite the sensors’ high nominal precision ($\pm 0.2\,^{\circ}$C and $\pm 0.4\,^{\circ}$C under solar radiation with speed $>7$\,km/h), occasional temperature anomalies, such as transient peaks, can arise due to sudden environmental variations like changes in sun exposure or airflow. For example, Figures \ref{fig:technical_val1}c and \ref{fig:technical_val1}d, which depict the temporal evolution of temperature and humidex during a thermal walk, reveal minor discrepancies between the two sensors on the same cart (indicated by blue and orange lines), particularly around temperature peaks. Averaging the readings from both sensors (black line) effectively mitigates these discrepancies, producing a cleaner and more reliable trajectory.

In the illustrated case, the average temperatures recorded by sensors 11 and 12 are $29.08 \pm 0.02\,^\circ\mathrm{C}$ and $28.75 \pm 0.02\,^\circ\mathrm{C}$, respectively, while the averaged temperature is $28.92 \pm 0.01\,^\circ\mathrm{C}$. Similarly, the average humidex values for sensors 11 and 12 are $33.73 \pm 0.03\,^\circ\mathrm{C}$ and $33.44 \pm 0.02\,^\circ\mathrm{C}$, respectively, with the combined average yielding $33.59 \pm 0.02\,^\circ\mathrm{C}$. This averaging procedure was also applied to the GPS coordinates, whereby latitude and longitude values from both sensors were averaged at each timestamp. As shown in Figures \ref{fig:technical_val_gps}, \ref{fig:technical_val_gps2}c, \ref{fig:technical_val_gps2}d, and \ref{fig:technical_val_gps2}e, the GPS positions reported by the two sensors generally exhibit strong agreement. Averaging these coordinates introduces negligible distortion to the spatial trajectories and, in fact, enhances geolocation accuracy in instances where minor divergence between the sensors’ GPS readings occurs, since they are primarily subject to satellite error rather than environmental variability. 

\section*{Usage Notes}
This dataset offers valuable opportunities for research in urban climatology, public health, and social sciences. Its integration of microclimatic sensor data, spatial and environmental descriptors, and thermal perception surveys allows both fine-grained analyses of urban heat exposure and comparative studies across neighborhoods, participant groups, and public outdoor space typologies. Potential applications include the validation of urban heat island models, the study of socio-demographic inequalities in thermal exposure, and the design of evidence-based adaptation strategies in climate-vulnerable areas.

The citizen science approach underlying the dataset also makes it relevant for participatory research and policy-making, providing a concrete example of how collaborative methodologies can generate high-quality scientific data while fostering community engagement. Although the dataset is specific to the Barcelona metropolitan area, its structure facilitates comparison with other urban contexts and encourages replication of the methodology elsewhere. Data quality protocols were systematically applied throughout the campaigns to ensure reliable sensor measurements and robust survey responses.

Beyond scientific research, the dataset may be used in education and training, serving as a resource for teaching data analysis, visualization, and participatory methods. Its open accessibility invites reuse and integration with complementary sources, supporting both academic inquiry and applied decision-making for more equitable and climate-resilient cities.

All datasets are openly available in the data repository \cite{Croniquesdata2025}. Raw geolocated microclimatic measurements obtained with the sensors are provided in the \textit{original\_trajectories} folder, while processed and cleaned trajectories--following the procedures described in this Data Descriptor--are available in the \textit{processed\_ trajectories} folder. Perceptual data from the thermal surveys can be accessed in the \textit{processed\_surveys} and \textit{aggregated\_surveys} folders. To ensure reproducibility, the repository \url{https://github.com/ferranlarroyaub/Croniques_de_la_Calor_2024}
 provides open Python code in Jupyter Notebooks, allowing users to replicate data processing steps, reproduce figures presented here, and extend the analysis and visualization.

\section*{Acknowledgements}
We thank all the organizations and individuals who actively participated in this research. The individual contributors have participated as co-researchers. They are children and adolescents attending summer camps, children and adolescents residing in shelter homes, primary school students (aged 10–11 years), secondary school students (aged 14–15 years), secondary school students receiving specialized educational support (aged 14–16 years), educators and teachers, mothers of toddlers, public library users, shopkeepers and older adult groups (women and mixed) participating in community activities. 

Their contributions were possible thanks to the  involvement of 14 collaborating community organizations across the five locations in the Barcelona metropolitan that facilitated the Heat Chronicles implementation in the territory: \emph{Fundació Comtal} (education and support for children, adolescents, and young people in situations of social vulnerability), \emph{Canòdrom} (Center for Digital and Democratic Innovation),\emph{Casal de Barri Congrés–Indians} (Community Center), \emph{Fundació MAIN} (Shared Schooling Unit – Specialized Education), \emph{Servei Residencial d’Acció Educativa Congrés–ISOM} (Shelter homes for children and adolescents), \emph{Unió de Botiguers} (Shopkeepers’ Union), \emph{Taula Comunitària del Barri del Congrés–Indians} (Community Assembly of the neigbhorhood), \emph{Biblioteca Josep Janés} (Public Library), \emph{Escola Pep Ventura} (Primary School), \emph{Escola Bressol Nova Fortuny} (Kindergarten), \emph{Procés Comunitari Intercultural} (Intercultural Community Process), \emph{Associació Educativa Itaca} (socio-educational organization promoting equal opportunities for children and young people), \emph{Institut La Ribera} (Secondary School), and \emph{Escola La Guàrdia} (Primary School). We specially thank Raquel Segura (Biblioteca Josep Janés), Mar Escarrabill (Canòdrom), Xavi Geis ( Generalitat de Catalunya) and Jaume Puigpinós (Taula del tercer sector) for their proactive contributions. We also thank team members that, in addition to author, have contributed coordinating and/or offering support to the campaigns: Maday Rivero, Iván Casanovas, Rebekah Breding, and Martín F. Díaz (OpenSystems group members), and  Ferran Capell, Roger Jové, Uriel Merino, Ulises Fabián Náger, Víctor Ramos, César Recacha, and Oriol Riera (students from the Faculty of Physics at the University of Barcelona). 

Heat Chronicles is partially funded by SENSE from the European Union’s Horizon 2020 Research and Innovation Program under Grant Agreement no. 101058507 and by OPUSH from the ERA-NET Urban Transformation Capacities (ENUTC) program under Grant Agreement no. 101003758. The study was also partially supported the Ministerio de Ciencia e Innovación and Agencia Estatal de Investigación MCIN/AEI/10.13039/501100011033 and by “ERDF A way of making Europe", grant number PID2022-140757NB-I00 and by MCIN/ AEI/10.13039/501100011033 and European Union NextGeneration EU/PRTR, grant number PCI2022-132996.

\section*{Authors' contributions}
Ferran Larroya - data acquisition - project conception - data validation – writing - proofreading. Isabelle Bonhoure - data acquisition - project conception - data validation – writing - proofreading. Femke Min - data acquisition - proofreading. Josep Perell\'o - data acquisition - project conception - writing - proofreading.

\section*{Competing interests}
The authors declare that they have no known competing financial interests or personal relationships that could have influenced the work reported in this paper.

%\clearpage

%\clearpage

%\clearpage

%\clearpage

%\clearpage

%\clearpage

%\clearpage

%\clearpage

\begin{table}[h]
\caption{\label{tab:city_neighborhood} {\bf Locations involved in the thermal walks.} Data collection was conducted in five locations within the Barcelona metropolitan area, including both cities and neighborhoods. Each location is referred to by the abbreviation (Ab.) indicated below.}
\begin{adjustbox}{width=\textwidth}
\begin{tabular}{llc}
\hline\hline
Neighborhood & City & Ab. \\ \hline
Sant Pere, Santa Caterina i la Ribera & Barcelona & SP 
\\ 
El Congr\'es i els Indians & Barcelona & CI \\ 
Collblanc-La Torrassa & L’Hospitalet de Llobregat & CT \\ 
- & Montcada i Reixac & MR \\ 
- & Sant Vicenç dels Horts & SV \\ 
 \hline
 \hline
\end{tabular}
\end{adjustbox}
\end{table}

\begin{table}[h]
\centering
\caption{\label{tab:partners}
{\bf Locations and partner institutions.} Overview of all participating institutions and community organizations that collaborated in the campaigns across the five locations in the Barcelona metropolitan area. The descriptions provided characterize each institution’s general mission. Abbreviations listed are used consistently in the manuscript to identify each location (cf. Table \ref{tab:city_neighborhood}).}
\begin{tabularx}{\textwidth}{l X}
\hline
\hline
Location & Partner Institution and Description \\
\hline
SP & Fundació Comtal \emph{(CSO focused on education and support for children, adolescents, and young people in situations of social vulnerability)} \\
\addlinespace
CI & Canòdrom – Center for Digital and Democratic Innovation\\
 & Casal de Barri Congrés-Indians \emph{(Community Center)}\\
 & Fundació MAIN \emph{(Shared Schooling Unit – Specialized Education)}\\
 & Servei Residencial d’Acció Educativa Congrés – ISOM \emph{(Shelter homes for children and adolescents)}\\
 & Unió de Botiguers \emph{(Shopkeepers’ Union)}\\
 & Taula Comunitària del Barri del Congrés-Indians \emph{(Community Assembly of the neighborhood)} \\
\addlinespace
CT & Biblioteca Josep Janés \emph{(Public Library)}\\
 & Escola Pep Ventura \emph{(Primary School)}\\
 & Escola Bressol Nova Fortuny \emph{(Kindergarten)}\\
 & Procés Comunitari Intercultural \emph{(Intercultural community process)}\\
 & Associació Educativa Itaca \emph{(CSO developing socio-educational projects to promote equal opportunities for children and young people)} \\
\addlinespace
MR & Institut La Ribera \emph{(Secondary School)} \\
\addlinespace
SV & Escola La Guàrdia \emph{(Primary School)} \\
 \hline
 \hline 
\end{tabularx}
\end{table}

\begin{table}[h]
\caption{\label{tab:city_participant} 
\textbf{Locations and participant groups.} 
Description of participant characteristics and the corresponding locations where participants were engaged. Abbreviations listed are used consistently in the manuscript to identify each location (cf. Table \ref{tab:city_neighborhood}).}
\begin{tabularx}{\textwidth}{l X}
\hline
\hline
Location & Participant group \\
\hline
SP, CT & Children and adolescents attending summer camps \\
\addlinespace
CI & Children and adolescents residing in shelter homes \\
CT, SV & Primary school students (aged 10–11 years) \\
MR & Secondary school students (aged 14–15 years) \\
CI & Secondary school students receiving specialized educational support (aged 14–16 years) \\
SP, CI, CT, MR, SV & Educators and teachers \\
CT & Mothers of toddlers \\
CT & Public library users \\
CI & Shopkeepers \\
CT, CI & Older adult groups (women and mixed) participating in community activities \\
\hline
\hline
\end{tabularx}
\end{table}

\begin{table}[h]
\centering
\caption{\label{tab:sociodem_all} \textbf{Sociodemographic statistics of all participants and thermal survey responses.} Summary of the sociodemographic profile based on the 439 unique participants and their thermal survey responses (TCVs, TSVs, and wTCVs). Percentages are calculated relative to the total number of participants or responses, as appropriate.}
\begin{adjustbox}{width=\textwidth}
\begin{tabular}{ccccc}
\hline \hline
& & Participants & TCVs/TSVs & wTCVs\\ \hline
\multirow{4}{4em}{Gender} & Woman & 225 (51\%) & 958 (51\%) & 738 (51\%) \\
& Man & 197 (45\%) & 837 (45\%) & 642 (45\%)\\
& Non-binary & 5 (1\%) & 21 (1\%) & 16 (1\%)\\
& I prefer not to say & 12 (3\%) & 51 (3\%) & 39 (3\%)\\ \hline
\multirow{5}{4em}{Age range} & Less than 12 & 158 (36\%) & 687 (37\%) & 532 (37\%) \\
& 13-15 & 181 (41\%) & 736 (39\%) & 555 (39\%) \\
& 16-24 & 46 (11\%) & 196 (11\%) & 153 (11\%) \\
& 25-54 & 32 (7\%) & 148 (8\%) & 117 (8\%)\\ 
& 55-84 & 22 (5\%) & 100 (5\%) & 78 (5\%) \\ \hline
\multirow{4}{10em}{Time spent each day in outdoor public spaces} & Less than 1h & 85 (20\%) & 379 (20\%) & 296 (20\%) \\
& 1h - 2h & 114 (26\%) & 478 (26\%) & 365 (26\%) \\
& 2h - 4h & 131 (30\%) & 552 (30\%) & 424 (30\%) \\
& More than 4h & 109 (24\%) & 458 (24\%) & 350 (24\%) \\ 
\hline 
\multirow{4}{8em}{Neighborhood knowledge level} & None or partial & 74 (17\%) & 329 (18\%) & 256 (18\%)\\
& Moderate & 90 (21\%) & 384 (20\%) & 296 (20\%) \\
& Good & 124 (28\%) & 526 (28\%) & 404 (28\%) \\
& Excellent & 151 (34\%) & 628 (34\%) & 479 (34\%) \\ 
\hline 
Total & & 439 & 1\,867 & 1\,435 \\ \hline \hline
\end{tabular}
\end{adjustbox}
\end{table}

\begin{table}[h]
\centering
\caption{\label{tab:sociodem_SP} \textbf{Sociodemographic statistics for \textit{Sant Pere, Santa Caterina i la Ribera} (Barcelona, SP).} Summary of the sociodemographic profile of participants in SP, based on the number of unique participants and their thermal survey responses (TCVs, TSVs, and wTCVs). Percentages are relative to the local total.}
\begin{adjustbox}{width=\textwidth}
\begin{tabular}{ccccc}
\hline \hline
& & Participants & TCVs/TSVs & wTCVs \\ \hline
\multirow{4}{4em}{Gender} & Woman & 53 (62\%) & 209 (65\%) & 159 (66\%) \\
& Man & 31 (37\%) & 109 (34\%) & 79 (32\%) \\
& Non-binary & 1 (1\%) & 5 (1\%) & 4 (2\%)\\
& I prefer not to say & 0 (0\%) & 0 (0\%) & 0 (0\%) \\ \hline
\multirow{5}{4em}{Age range} & Less than 12 & 29 (34\%) & 106 (33\%) & 79 (33\%) \\
& 13-15 & 31 (37\%) & 120 (37\%) & 89 (37\%)\\
& 16-24 & 18 (21\%) & 70 (22\%) & 54 (22\%) \\
& 25-54 & 7 (8\%) & 27 (8\%) & 20 (8\%) \\ 
& 55-84 & 0 (0\%) & 0 (0\%) & 0 (0\%) \\ \hline
\multirow{4}{10em}{Time spent each day in outdoor public spaces} & Less than 1h & 14 (17\%) & 54 (17\%) & 42 (17\%) \\
& 1h - 2h & 22 (26\%) & 82 (25\%) & 60 (25\%) \\
& 2h - 4h & 24 (28\%) & 90 (28\%) & 68 (28\%)\\
& More than 4h & 25 (29\%) & 97 (30\%) & 72 (30\%) \\ 
\hline 
\multirow{4}{8em}{Neighborhood knowledge level} & None or partial & 15 (18\%) & 57 (17\%) & 42 (17\%) \\
& Moderate & 20 (24\%) & 73 (23\%) & 54 (22\%) \\
& Good & 25 (29\%) & 93 (29\%) & 70 (29\%) \\
& Excellent & 25 (29\%) & 100 (31\%) & 76 (32\%) \\ 
\hline 
Total & & 85 & 323 & 242\\ \hline \hline
\end{tabular}
\end{adjustbox}
\end{table}

\begin{table}[h]
\centering
\caption{\label{tab:sociodem_CI} \textbf{Sociodemographic statistics for \textit{El Congrés i els Indians} (Barcelona, CI).} Summary of the sociodemographic profile of participants in CI, based on the number of unique participants and their thermal survey responses (TCVs, TSVs, and wTCVs). Percentages are relative to the local total.}
\begin{adjustbox}{width=\textwidth}
\begin{tabular}{ccccc}
\hline \hline
& & Participants & TCVs/TSVs & wTCVs \\ \hline
\multirow{4}{4em}{Gender} & Woman & 24 (51\%) & 107 (45\%) & 84 (45\%) \\
& Man & 21 (45\%) & 121 (51\%) & 100 (52\%) \\
& Non-binary & 1 (2\%) & 5 (2\%) & 4 (2\%)\\
& I prefer not to say & 1 (2\%) & 5 (2\%) & 4 (2\%)\\ \hline
\multirow{5}{4em}{Age range} & Less than 12 & 8 (17\%) & 40 (17\%) & 32 (17\%) \\
& 13-15 & 9 (19\%) & 58 (24\%) & 49 (25\%) \\
& 16-24 & 10 (21\%) & 52 (22\%) & 43 (22\%) \\
& 25-54 & 7 (15\%) & 35 (15\%) & 28 (15\%) \\ 
& 55-84 & 13 (28\%) & 53 (22\%) & 40 (21\%) \\ \hline
\multirow{4}{10em}{Time spent each day in outdoor public spaces} & Less than 1h & 10 (21\%) & 48 (20\%) & 38 (20\%) \\
& 1h - 2h & 13 (28\%) & 58 (25\%) & 45 (23\%) \\
& 2h - 4h & 14 (30\%) & 79 (33\%) & 65 (34\%)\\
& More than 4h & 10 (21\%) & 53 (22\%) & 44 (23\%) \\ 
\hline 
\multirow{4}{8em}{Neighborhood knowledge level} & None or partial & 16 (34\%) & 85 (36\%) & 69 (36\%) \\
& Moderate & 12 (26\%) & 57 (24\%) & 45 (23\%) \\
& Good & 8 (17\%) & 42 (17\%) & 34 (18\%) \\
& Excellent & 11 (23\%) & 54 (23\%) & 44 (23\%) \\ 
\hline 
Total & & 47 & 238 & 192 \\ \hline \hline
\end{tabular}
\end{adjustbox}
\end{table}

\begin{table}[h]
\centering
\caption{\label{tab:sociodem_CT} \textbf{Sociodemographic statistics for \textit{Collblanc-La Torrassa} (L'Hospitalet de Llobregat, CT).} Summary of the sociodemographic profile of participants in CT, based on the number of unique participants and their thermal survey responses (TCVs, TSVs, and wTCVs). Percentages are relative to the local total.}
\begin{adjustbox}{width=\textwidth}
\begin{tabular}{ccccc}
\hline \hline
& & Participants & TCVs/TSVs & wTCVs \\ \hline
\multirow{4}{4em}{Gender} & Woman & 81 (50\%) & 368 (50\%) & 287 (51\%) \\
& Man & 84 (49\%) & 357 (49\%) & 274 (48\%) \\
& Non-binary & 0 (0\%) & 0 (0\%) & 0 (0\%) \\
& I prefer not to say & 2 (1\%) & 8 (1\%) & 6 (1\%) \\ \hline
\multirow{5}{4em}{Age range} & Less than 12 & 104 (62\%) & 456 (62\%) & 353 (62\%) \\
& 13-15 & 32 (19\%) & 131 (18\%) & 99 (18\%) \\
& 16-24 & 12 (7\%) & 48 (7\%) & 36 (7\%) \\
& 25-54 & 10 (6\%) & 51 (7\%) & 41 (7\%) \\ 
& 55-84 & 9 (6\%) & 47 (6\%) & 38 (6\%) \\ \hline
\multirow{4}{10em}{Time spent each day in outdoor public spaces} & Less than 1h & 36 (22\%) & 170 (23\%) & 134 (24\%) \\
& 1h - 2h & 48 (28\%) & 215 (29\%) & 168 (30\%)\\
& 2h - 4h & 47 (28\%) & 191 (26\%) & 144 (25\%)\\
& More than 4h & 36 (22\%) & 157 (22\%) & 121 (21\%) \\ 
\hline 
\multirow{4}{8em}{Neighborhood knowledge level} & None or partial & 28 (17\%) & 122 (17\%) & 95 (17\%) \\
& Moderate & 42 (25\%) & 187 (25\%) & 145 (25\%) \\
& Good & 55 (33\%) & 241 (33\%) & 186 (33\%) \\
& Excellent & 42 (25\%) & 183 (25\%) & 141 (25\%) \\ 
\hline 
Total & & 167 & 733 & 567 \\ \hline \hline
\end{tabular}
\end{adjustbox}
\end{table}

\begin{table}[h]
\centering
\caption{\label{tab:sociodem_MR} \textbf{Sociodemographic statistics for \textit{Montcada i Reixac} (MR).} Summary of the sociodemographic profile of participants in MR, based on the number of unique participants and their thermal survey responses (TCVs, TSVs, and wTCVs). Percentages are relative to the local total.}
\begin{adjustbox}{width=\textwidth}
\begin{tabular}{ccccc}
\hline \hline
& & Participants & TCVs/TSVs & wTCVs \\ \hline
\multirow{4}{4em}{Gender} & Woman & 59 (48\%) & 234 (48\%) & 176 (49\%) \\
& Man & 52 (43\%) & 205 (43\%) & 153 (42\%) \\
& Non-binary & 3 (2\%) & 11 (2\%) & 8 (2\%) \\
& I prefer not to say & 8 (7\%) & 33 (7\%) & 25 (7\%)\\ \hline
\multirow{5}{4em}{Age range} & Less than 12 & 0 (0\%) & 0 (0\%) & 0 (0\%) \\
& 13-15 & 109 (89\%) & 427 (89\%) & 318 (88\%) \\
& 16-24 & 6 (5\%) & 26 (5\%) & 20 (6\%) \\
& 25-54 & 7 (6\%) & 30 (6\%) & 24 (6\%)\\ 
& 55-84 & 0 (0\%) & 0 (0\%) & 0 (0\%) \\ \hline
\multirow{4}{10em}{Time spent each day in outdoor public spaces} & Less than 1h & 18 (15\%) & 72 (15\%) & 54 (15\%) \\
& 1h - 2h & 28 (23\%) & 108 (22\%) & 80 (22\%) \\
& 2h - 4h & 42 (34\%) & 172 (36\%) & 131 (36\%) \\
& More than 4h & 34 (28\%) & 131 (27\%) & 97 (27\%) \\ 
\hline 
\multirow{4}{8em}{Neighborhood knowledge level} & None or partial & 13 (11\%) & 55 (11\%) & 42 (11\%) \\
& Moderate & 12 (10\%) & 47 (10\%) & 36 (10\%) \\
& Good & 30 (24\%) & 120 (25\%) & 90 (25\%) \\
& Excellent & 67 (55\%) & 261 (54\%) & 194 (54\%) \\ 
\hline 
Total & & 122 & 483 & 362 \\ \hline \hline
\end{tabular}
\end{adjustbox}
\end{table}

\begin{table}[h]
\centering
\caption{\label{tab:sociodem_SV} \textbf{Sociodemographic statistics for \textit{Sant Vicenç dels Horts} (SV).} Summary of the sociodemographic profile of participants in SV, based on the number of unique participants and their thermal survey responses (TCVs, TSVs, and wTCVs). Percentages are relative to the local total.}
\begin{adjustbox}{width=\textwidth}
\begin{tabular}{ccccc}
\hline \hline
& & Participants & TCVs/TSVs & wTCVs \\ \hline
\multirow{4}{4em}{Gender} & Woman & 8 (44\%) & 40 (44\%) & 32 (44\%) \\
& Man & 9 (50\%) & 45 (50\%) & 36 (50\%) \\
& Non-binary & 0 (0\%) & 0 (0\%) & 0 (0\%) \\
& I prefer not to say & 1 (6\%) & 5 (6\%) & 4 (6\%) \\ \hline
\multirow{5}{4em}{Age range} & Less than 12 & 17 (94\%) & 85 (94\%) & 68 (94\%) \\
& 13-15 & 0 (0\%) & 0 (0\%) & 0 (0\%) \\
& 16-24 & 0 (0\%) & 0 (0\%) & 0 (0\%) \\
& 25-54 & 1 (6\%) & 5 (6\%) & 4 (6\%) \\ 
& 55-84 & 0 (0\%) & 0 (0\%) & 0 (0\%) \\ \hline
\multirow{4}{10em}{Time spent each day in outdoor public spaces} & Less than 1h & 7 (39\%) & 35 (39\%) & 28 (39\%) \\
& 1h - 2h & 3 (17\%) & 15 (17\%) & 12 (17\%)\\
& 2h - 4h & 4 (22\%) & 20 (22\%) & 16 (22\%) \\
& More than 4h & 4 (22\%) & 20 (22\%) & 16 (22\%) \\ 
\hline 
\multirow{4}{8em}{Neighborhood knowledge level} & None or partial & 2 (11\%) & 10 (11\%) & 8 (11\%) \\
& Moderate & 4 (22\%) & 20 (22\%) & 16 (22\%) \\
& Good & 6 (33\%) & 30 (33\%) & 24 (33\%) \\
& Excellent & 6 (33\%) & 30 (33\%) & 24 (33\%) \\ 
\hline 
Total & & 18 & 90 & 72\\ \hline \hline
\end{tabular}
\end{adjustbox}
\end{table}

\begin{table}[h]
\centering
\caption{\label{tab:sociodem_place} \textbf{Summary statistics for the five locations.} The table reports the number of individual thermal walk trajectories (groups), participants, surveyed sites (where thermal surveys were conducted), and the number of self-reported entries, including TCVs, TSVs, and wTCVs.}
\begin{adjustbox}{width=\textwidth}
\begin{tabular}{cccccc}
\hline \hline
Place & Groups (trajectories) & Participants & Sites & TCVs/TSVs & wTCVs \\ \hline
SV & 3 (6\%) & 18 (4\%) & 15 (7\%) & 90 (5\%) & 72 (5\%) \\
CT & 16 (33\%) & 167 (38\%) & 71 (34\%) & 733 (39\%) & 567 (40\%) \\
MR & 12 (25\%) & 122 (28\%) & 48 (23\%) & 483 (26\%) & 362 (25\%) \\
SP & 10 (21\%) & 84 (19\%) & 40 (19\%) & 323 (17\%) & 242 (17\%) \\
CI & 7 (15\%) & 47 (11\%) & 36 (17\%) & 238 (13\%) & 192 (13\%)\\ 
\hline 
Total & 48 & 439 & 210 & 1867 & 1435 \\ \hline \hline
\end{tabular}
\end{adjustbox}
\end{table}

\begin{table}[h]
\caption{\label{tab:space_categories} 
\textbf{Public site categories.} 
During the initial session with community participants, several sites of interest were identified for thermal surveys. Each location was marked on a map and subsequently classified according to the categories listed below.}
\begin{tabularx}{\textwidth}{l l c X}
\hline\hline
Macro category & Site category & Code & Description \\
\hline
Basic Needs
& Transit spaces & TS & Streets and squares used for daily circulation \\
& Health & H & Primary care centers, hospitals, day centers, pharmacies \\
& Education & E & Nursery schools, primary schools, high schools \\
& Transport & T & Metro, train, or bus stations; bus and tram stops \\
& Markets & M & Municipal markets \\
\hline
Wellbeing 
& Rest places & RP & Parks, green areas, fountains \\
and Free Time & Meeting places & MP & Squares, playgrounds, pedestrian streets \\
& Sports facilities & SE & Sports centers, swimming pools, tracks \\
& Cultural facilities & CF & Libraries, civic centers, museums \\
\hline\hline
\end{tabularx}
\end{table}

\begin{table}[h]
\caption{\label{tab:driver_sheet} \textbf{Driver’s sheet.} The driver was responsible for managing the group during the thermal walk and for completing the initial section of the table at the starting point, which included information about the group. At each surveyed site, the driver recorded environmental observations such as the number of nearby individuals, ambient noise levels, and the presence and intensity of unpleasant odors. Additionally, the driver documented the arrival time, the time of the thermal surveys, and the departure time at each site.}
\begin{adjustbox}{width=\textwidth}
\begin{tabular}{llllll}
\hline \hline

\multicolumn{3}{l|}{\textbf{DRIVER'S NAME:}} & cart NUMBER: \\ 
\multicolumn{3}{l|}{\textbf{PLACE:}} & DATE: \\ \hline
\multicolumn{6}{l}{\textbf{STARTING POINT. GROUP PROFILE}} \\ 
\multicolumn{6}{l}{\textbf{Arrival time: \_\_h\_\_min}} \\
\multicolumn{6}{l}{\textbf{1. Number of participants in this group: }} \\ \hline
\multicolumn{6}{l}{\textbf{2. Indicate the age range(s) of majority (in years):}} \\
Less than 10 & 10 - 12 & 13 - 15 & 16 - 18 & 19 - 24 & 25 - 34 \\
35 - 44 & 45 - 54 & 55 - 64 & 65 - 74 & 75 - 84 & More than 85 \\ \hline
\multicolumn{6}{l}{\textbf{3. How are the participants dressed? (influences heat perception)}} \\
\multirow{2}{6em}{All with long pants/skirts and long sleeves} & \multirow{2}{6em}{Most with long pants/skirts and long sleeves} & \multirow{2}{6em}{Half ``long'' and half ``short''} & \multirow{2}{6em}{Most with long pants/skirts and short sleeves} & \multirow{2}{6em}{All with short pants/skirts and short sleeves} & \\ 
 & & & & & \\
 & & & & & \\
 & & & & & \\
 & & & & & \\
\hline
\hline
& & & & & \\
\multicolumn{6}{l}{\textbf{STARTING POINT. SURVEY 1} Site code:} \\ 
\multicolumn{6}{l}{\textbf{SURVEY 1 time: \_\_h\_\_min // Departure time: \_\_h\_\_min}} \\ \hline
\multicolumn{6}{l}{\multirow{2}{50em}{\textbf{1. Within a radius of approximately 10 meters (about 10 steps), how many people are there? (excluding the people in your group)}}} \\
& & & & & \\
Nobody & 1 - 3 people & 4 - 10 people & 10 - 20 people & More than 20 people & \\ \hline
\multicolumn{6}{l}{\textbf{2. What is the noise level you perceive?}} \\
No noise or very low noise & Moderate noise & Loud noise & & & \\ \hline
\multicolumn{6}{l}{\textbf{3. What is the intensity of bad smell that you perceive?}} \\
No bad smell or very weak & Moderate bad smell & Strong bad smell & & & \\ \hline \hline
& & & & & \\ 
\multicolumn{6}{l}{\textbf{FIRST STOP. SURVEY 2} Site code: } \\ 
\multicolumn{6}{l}{\textbf{Arrival time: \_\_h \_\_min: // SURVEY 2 time: \_\_h\_\_min // Departure time: \_\_h\_\_min}} \\ \hline
\multicolumn{6}{l}{\multirow{2}{50em}{\textbf{4. Within a radius of approximately 10 meters (about 10 steps), how many people are there? (excluding the people in your group)}}} \\
& & & & & \\
Nobody & 1 - 3 people & 4 - 10 people & 10 - 20 people & More than 20 people & \\ \hline
\multicolumn{6}{l}{\textbf{5. What is the noise level you perceive?}} \\
No noise or very low noise & Moderate noise & Loud noise & & & \\ \hline
\multicolumn{6}{l}{\textbf{6. What is the intensity of bad smell that you perceive?}} \\
No bad smell or very weak & Moderate bad smell & Strong bad smell & & & \\ \hline \hline
\end{tabular}
\end{adjustbox}
\end{table}

\begin{table}[h]
\caption{\label{tab:sociodem_english} \textbf{Sociodemographic profile of participants before a thermal walk.} Participants completed a questionnaire covering four sociodemographic aspects: gender, age, neighborhood knowledge, and time spent in outdoor public spaces. Figure~\ref{fig:votes}b shows the physical form filled out by participants (in Catalan). Each completed form (ballot) is linked to the participant's unique ID.}
\begin{adjustbox}{width=\textwidth}
\begin{tabular}{llllll}
\hline \hline
\multicolumn{5}{l|}{\textbf{Your profile}} & ID: XXX \\ \hline
\multicolumn{6}{l}{\textbf{1. Mark the gender identity with which you identify most}} \\
Woman & Man & Non-binary & I prefer not to say & & \\ \hline
\multicolumn{6}{l}{\textbf{2. Please indicate your age (in years)}} \\
Less than 10 & 10 - 12 & 13 - 15 & 16 - 18 & 19 - 24 & 25 - 34 \\
35 - 44 & 45 - 54 & 55 - 64 & 65 - 74 & 75 - 84 & More than 85 \\ \hline
\multicolumn{6}{l}{\textbf{3. How would you define your level of knowledge of this neighborhood?} } \\
No knowledge & Partial knowledge & Moderate knowledge & Good knowledge & Excellent knowledge & \\ \hline
\multicolumn{6}{l}{\textbf{4. On average, how many hours per day do you spend in outdoor public spaces (e.g., streets, squares, parks)?}} \\
Less than 30 minutes & 30 minutes – 1 hour & 1 hour – 2 hours & 2 hours – 4 hours & more than 4 hours & \\
\hline \hline
\end{tabular}
\end{adjustbox}
\end{table}

\begin{table}[h]
\caption{\label{tab:vote1_english} \textbf{Thermal survey: Ballot 1.} At the starting point and before the thermal walk, participants responded to four questions related to their perception of heat, including thermal comfort, thermal sensation, wind, and sun exposure. Figure~\ref{fig:votes}c shows the physical form completed by participants (in Catalan). Each completed ballot is associated with the participant's unique ID and the code corresponding to the site where the survey was conducted.}
\begin{adjustbox}{width=\textwidth}
\begin{tabular}{llllllll}
\hline \hline
\multicolumn{4}{l|}{\textbf{BALLOT 1}} & \multicolumn{2}{l|}{Site code: ---} & ID: XXX \\ \hline
\multicolumn{7}{l}{\textbf{1. Right now, how are you feeling? In terms of thermal comfort, I find myself...}} \\
1 very uncomfortable & 2 uncomfortable & 3 slightly uncomfortable & 4 neutral & 5 slightly comfortable & 6 comfortable & 7 very comfortable \\ \hline
\multicolumn{7}{l}{\textbf{2. Right now, what is the space like according to your thermal sensation?}} \\
1 Cool & 2 Slightly cool & 3 Neutral & 4 Slightly warm & 5 Warm & 6 Hot & 7 Very hot \\ \hline
\multicolumn{7}{l}{\textbf{3. Right now, what is the wind blowing?} } \\
1 It's not windy & 2 A very light wind & 3 A light wind & 4 A moderate wind & 5 A strong wind & & \\ \hline
\multicolumn{7}{l}{\textbf{4. Right now, I am...}} \\
1 in full sun & 2 in a mixture of sun and shadow & 3 in full shade & & & & \\
\hline \hline
\end{tabular}
\end{adjustbox}
\end{table}

\begin{table}[h]

\caption{\label{tab:vote2_english} \textbf{Thermal survey: Ballot 2.} At each site along the route, the participants responded to five questions concerning their heat perception, including thermal comfort while walking, overall thermal comfort, thermal sensation, wind perception, and sun exposure. Figure~\ref{fig:votes}d shows the physical form completed by participants (in Catalan). Each completed ballot is associated with the participant's unique ID and the code corresponding to the site where the survey was conducted. From ballot 2 onward, the thermal survey questions remained consistent at all stops in each site along the route.}
\begin{adjustbox}{width=\textwidth}
\begin{tabular}{llllllll}
\hline \hline
\multicolumn{4}{l|}{\textbf{BALLOT 2}} & \multicolumn{2}{l|}{Site code: ---} & ID: XXX \\ \hline

\multicolumn{7}{l}{\textbf{1. How did you feel while walking? In terms of thermal comfort, I have found myself...}} \\
1 very uncomfortable & 2 uncomfortable & 3 slightly uncomfortable & 4 neutral & 5 slightly comfortable & 6 comfortable & 7 very comfortable \\ \hline

\multicolumn{7}{l}{\textbf{2. Right now, how are you feeling? In terms of thermal comfort, I find myself...}} \\
1 very uncomfortable & 2 uncomfortable & 3 slightly uncomfortable & 4 neutral & 5 slightly comfortable & 6 comfortable & 7 very comfortable \\ \hline
\multicolumn{7}{l}{\textbf{3. Right now, what is the space like according to your thermal sensation?}} \\
1 Cool & 2 Slightly cool & 3 Neutral & 4 Slightly warm & 5 Warm & 6 Hot & 7 Very hot \\ \hline
\multicolumn{7}{l}{\textbf{4. Right now, what is the wind blowing?} } \\
1 It's not windy & 2 A very light wind & 3 A light wind & 4 A moderate wind & 5 A strong wind & & \\ \hline
\multicolumn{7}{l}{\textbf{5. Right now, I am...}} \\
1 in full sun & 2 in a mixture of sun and shadow & 3 in full shade & & & & \\
\hline \hline
\end{tabular}
\end{adjustbox}
\end{table}

\begin{table}[h]
\caption{\label{tab:columns_raw_trajectories} \textbf{Columns of the \texttt{.csv} files for each individual trajectory.} Description of the ten retained columns in the \texttt{.csv} files generated by the MeteoTracker sensors, after discarding empty or irrelevant columns.}
\begin{adjustbox}{width=\textwidth}
\begin{tabular}{ll}
\hline\hline
Column & Description \\ \hline
Time & Local Timestamp of each record (YYYY:MM:DD HH:MM:SS+HH:SS) \\ 
Lat & Latitude coordinate of each record in degrees \\ 
Lon & Longitude coordinate of each record in degrees\\ 
Temp[\textdegree C] & Air temperature in degrees \\ 
Hum[\%] & Relative humidity in \% \\ 
Alt[m] & Altitude above the sea level in meters \\ 
Press[mbar] & Preassure in mbar\\ 
DP[\textdegree C] & Dew-point temperature in degrees \\ 
HDX[\textdegree C] & Humidex (humidity index) in degrees \\ 
Speed[km/h] & Speed in km/h \\ 
 \hline
 \hline
\end{tabular}
\end{adjustbox}
\end{table}

\begin{table}[h]
\caption{\label{tab:example_trajectory_file} \textbf{Example of a raw \texttt{.csv} file display of an individual trajectory.} Sample of raw data collected using a MeteoTracker sensor, shown after removing empty and irrelevant columns (see Table~\ref{tab:columns_raw_trajectories}). The data corresponds to a group participating in the thermal walk in the SP neighborhood, on July 10, 2024. It contains 4\,879 records spanning from 2:51~pm to 4:46~pm.}
\begin{adjustbox}{width=\textwidth}
\begin{tabular}{ccccccccccc}
\hline\hline
& Time & Lat & Lon & Temp[\textdegree C] & Hum[\%] & Alt[m] & Press[mbar] & DP[\textdegree C] & HDX[\textdegree C] & Speed[km/h] \\ \hline
0 & 2024-07-10T14:51:08+02:00 & 41.389610 & 2.178895 & 28.5 & 69 & 13 & 1015.0 & 22.3 & 39.4 & 0 \\ 
1 & 2024-07-10T14:51:10+02:00 & 41.389610 & 2.178895 & 28.5 & 69 & 13 & 1015.0 & 22.3 & 39.4 & 0 \\ 
2 & 2024-07-10T14:51:11+02:00 & 41.389610 & 2.178895 & 28.5 & 69 & 13 & 1015.0 & 22.3 & 39.4 & 0 \\ 
...&...&...&...&...&...&...&...&...&...&...\\
4876 & 2024-07-10T16:46:18+02:00 & 41.389612 & 2.178999 & 30.0 & 64 & 20 & 1015.0 & 22.8 & 41.3 & 0 \\
4877 & 2024-07-10T16:46:20+02:00 & 41.389612 & 2.178999 & 30.0 & 64 & 19 & 1015.0 & 22.8 & 41.4 & 0 \\ 
4878 & 2024-07-10T16:46:21+02:00 & 41.389612 & 2.178999 & 30.0 & 64 & 19 & 1015.0 & 22.8 & 41.4 &
0 \\
 \hline
 \hline
\end{tabular}
\end{adjustbox}
\end{table}

\begin{table}[h]
\caption{\label{tab:calibration_averages} \textbf{Indoor and outdoor validation tests.}
Mean temperature ($T$) and relative humidity ($RH$) values obtained during the indoor and outdoor tests shown in Figures~\ref{fig:technical_val_calibration3}, and \ref{fig:technical_val_calibration4}. Results include the mean and the standard error of the mean. Sensor readings within each test mostly fall within the documented precision range of the device.}
\begin{adjustbox}{width=\textwidth}
\begin{tabular}{c|c|c|c}
\hline \hline
\multicolumn{1}{l}{} & & $\langle T\rangle \pm \delta \, T (^{\circ}C)$ & $\langle RH\rangle \pm \delta \, RH (\%)$ \\ \hline
\multirow{3}{*}{Test 1 (indoor)} & Sensor 2 & $21.686\pm 0.003$ & $55\pm 0$ \\ & Sensor 3 & $21.739\pm 0.003$ & $54\pm 0$ \\ & Sensor 4 & $21.739\pm 0.001$ & $54 \pm 0.01$ \\ \hline
\multirow{3}{*}{Test 2 (indoor)} & Sensor 11 & $20.309\pm 0.002$ & $57\pm 0$ \\ & Sensor 1 & $20.348\pm 0.004$ & $57\pm 0.04$ \\ & Sensor 5 & $20.355\pm 0.004$ & $57\pm 0.01$ \\ \hline
\multirow{2}{*}{Test 3 (indoor)} & Sensor 6 & $21.440\pm 0.003$ & $56\pm 0.02$ \\ & Sensor 9 & $21.584\pm 0.003 $ & $57\pm 0.05$ \\ \hline
\multirow{3}{*}{Test 4 (indoor)} & Sensor 20 & $20.283\pm 0.003$ & $57\pm 0.03$ \\ & Sensor 10 & $20.313\pm 0.004$ & $58\pm 0.01$ \\ & Sensor 7 & $20.377\pm 0.004$ & $57\pm 0.02$ \\ \hline
\multirow{2}{*}{Test 5 (indoor)} & Sensor 12 & $21.364\pm 0.005$ & $58\pm 0.03$ \\ & Sensor 13 & $21.370\pm 0.003$ & $57\pm 0.03$ \\ \hline
\multirow{2}{*}{Test 6 (indoor)} & Sensor 14 & $22.001\pm 0.002$ & $54\pm 0.02$\\
& Sensor 15 & $22.117\pm 0.003$ & $54\pm 0.01$ \\ \hline 
\multirow{2}{*}{Test 7 (indoor)} & Sensor 8 & $22.102\pm 0.002$ & $55\pm 0$ \\ & Sensor 16 & $22.199\pm 0.002$ & $54\pm 0$ \\ \hline
\multirow{2}{*}{Test 8 (indoor)} & Sensor 7 & $21.807\pm 0.003$ & $53\pm 0.04$ \\ & Sensor 17 & $21.800\pm 0.000$ & $54\pm 0.04$ \\ \hline
\multirow{2}{*}{Test 9 (indoor)} & Sensor 18 & $21.493\pm 0.003$ & $57\pm 0.03$\\ & Sensor 19 & $21.519\pm 0.003$ & $57\pm 0.03$ \\ \hline
\multirow{4}{*}{Test 1 (outdoor)} & Sensor 19 & $18.47\pm 0.02$ & $47\pm 0.04$ \\ & Sensor 1 & $18.29\pm 0.02$ & $48\pm 0.05$ \\ & Sensor 3 & $18.44\pm 0.02$ & $46\pm 0.04$ \\ & Sensor 17 & $18.74\pm 0.02$ & $45\pm 0.04$ \\ \hline
\multirow{3}{*}{Test 2 (outdoor)} & Sensor 3 & $23.17\pm 0.04$ & $35\pm 0.1$ \\ & Sensor 10 & $23.15\pm 0.03$ & $36\pm 0.1$ \\ & Sensor 18 & $23.49\pm 0.02$ & $34\pm 0.1$ \\ \hline
\multirow{3}{*}{Test 3 (outdoor)} & Sensor 8 & $24.31\pm 0.02$ & $48\pm 0.1$ \\ & Sensor 15 & $23.38\pm 0.02$ & $47\pm 0.1$ \\ & Sensor 18 & $23.08\pm 0.02$ & $49\pm 0.1$ \\ \hline \hline
\end{tabular}
\end{adjustbox}
\end{table}

\end{document}